\documentclass[review,12pt]{elsarticle}
\usepackage{float,appendix,multicol}
\usepackage[utf8]{inputenc}    
\usepackage[T1]{fontenc} 
\usepackage{amsmath,amssymb, mathtools,mathrsfs,stmaryrd,titletoc}

\usepackage[numbers]{natbib}
\usepackage[retainorgcmds]{IEEEtrantools}
\usepackage[usenames]{color}
\usepackage{tabularx}
\usepackage{booktabs}
\usepackage{psfrag}
\usepackage[font=small,labelfont=md]{caption,subfig}
\usepackage{multirow}
\usepackage[T1]{fontenc} 
\usepackage[bookmarks=true,colorlinks=true,linkcolor=blue,citecolor=red]{hyperref}
\usepackage{float}         
\usepackage{listings}      
\usepackage{algorithm}
\usepackage{algorithmicx}
\usepackage{algpseudocode}
\usepackage{verbatim}
\usepackage{numprint}

\usepackage{silence}
\usepackage{siunitx}
\usepackage[norefs,nocites,ignoreunlbld]{refcheck} 
\usepackage[activate={true,nocompatibility},final,tracking=true,kerning=true,spacing=true,factor=1100,stretch=10,shrink=10]{microtype}

\usepackage[capitalise]{cleveref} 
\crefname{figure}{Fig.}{Figs.}
\crefname{equation}{Eq.}{Eqs.}
\crefname{section}{Section}{Sections}
\usepackage[textsize=tiny]{todonotes}
\usepackage{nicefrac} 
\usepackage{setspace}
\usepackage{lineno}  
\usepackage{totcount} 
\usepackage[figure,table]{totalcount}
\usepackage{blkarray, bigstrut} 
\usepackage{setspace}
\usepackage{tikz}
\usetikzlibrary{arrows,decorations.pathmorphing,decorations.pathreplacing,backgrounds,positioning,fit,matrix,math,shapes.misc}
\tikzset{cross/.style={cross out, draw=black, minimum size=2* (#1-\pgflinewidth), inner sep=0pt, outer sep=0pt}, cross/.default={1pt}}
\usepackage{wasysym}
\usepackage{gensymb} 
\usepackage[section]{placeins} 

\usepackage{xspace}

\newcommand{\ie}{i.e.,\xspace}

\newcommand{\trace}{\mathrm{tr}}
\newcommand{\bfsigma}{\boldsymbol{\sigma}}
\newcommand{\bfepsilon}{\boldsymbol{\varepsilon}}

\newcommand{\dd}{\mathrm{d}}
\newcommand*{\vd}[3][]{\dfrac{\mathrm{d}^{#1} #2}{\mathrm{d} #3^{#1} }}                          
\newcommand*{\pd}[3][]{\dfrac{\partial^{#1} #2}{\partial #3^{#1} }}

\usepackage{setspace}
\usepackage[many]{tcolorbox}
\tcbuselibrary{skins,breakable}
\definecolor{mycolor}{rgb}{0.122, 0.435, 0.698}
\newtcolorbox[auto counter,crefname={Box}{Boxes}]{MyBox}[2][]{%
  colback=gray!5!white,colframe=red!75!black,fonttitle=\bfseries,before skip=20pt plus
  2pt,after skip=20pt,
  title=Box~\thetcbcounter: #2, #1}

\newcounter{remark}[section]

\definecolor{Sun}{rgb}{0.164,0.126,0.322}
\definecolor{Green}{rgb}{0,0.300,0.300}
\definecolor{Red}{rgb}{0.4,0,0}
\definecolor{Grey}{RGB}{105,105,105}
\definecolor{White}{rgb}{1,1,1}

\algblockdefx{Start}{End}[1]{\textbf{#1}}{\textbf{end}}

\DeclareTextCommand{\nobreakspace}{T1}{\leavevmode\nobreak\ }

\graphicspath{{./}} 

\makeatletter
\def\@author#1{\g@addto@macro\elsauthors{\normalsize%
    \def\baselinestretch{1}%
    \upshape\authorsep#1\unskip\textsuperscript{%
      \ifx\@fnmark\@empty\else\unskip\sep\@fnmark\let\sep=,\fi
      \ifx\@corref\@empty\else\unskip\sep\@corref\let\sep=,\fi
      }%
    \def\authorsep{\unskip,\space}%
    \global\let\@fnmark\@empty
    \global\let\@corref\@empty  
    \global\let\sep\@empty}%
    \@eadauthor={#1}
}
\makeatother
\usepackage{longtable} 
\usepackage{lipsum} 
\usepackage[percent]{overpic}
\usepackage[T1]{fontenc}
\usepackage{lmodern}
\usepackage[margin=2.2cm]{geometry}
\biboptions{numbers,sort&compress}

\journal{International Journal of Hydrogen Energy}

\begin{document}
\begin{frontmatter}
\title{Computational predictions of weld structural integrity in hydrogen transport pipelines}

\author[Oxf,ICL]{Tushar Kanti Mandal} 

\author[2]{Jonathan Parker} 
\author[2]{Michael Gagliano} 
\author[Oxf,ICL]{Emilio Mart\'{\i}nez-Pa\~neda\corref{cor1}} 
\ead{emilio.martinez-paneda@eng.ox.ac.uk}

\address[Oxf]{Department of Engineering Science, University of Oxford, Oxford OX1 3PJ, UK}
\address[2]{Electric Power Research Institute, 3420 Hillview Avenue, Palo Alto, CA 94304, USA}
\address[ICL]{Department of Civil and Environmental Engineering, Imperial College London, London SW7 2AZ, UK}

\cortext[cor1]{Corresponding author.}

\begin{abstract}
We combine welding process modelling with deformation-diffusion-fracture (embrittlement) simulations to predict failures in hydrogen transport pipelines. The focus is on the structural integrity of seam welds, as these are often the locations most susceptible to damage in gas transport infrastructure. Finite element analyses are conducted to showcase the ability of the model to predict cracking in pipeline steels exposed to hydrogen-containing environments. The validated model is then employed to quantify critical H$_2$ fracture pressures. The coupled, phase field-based simulations conducted provide insight into the role of existing defects, microstructural heterogeneity, and residual stresses. We find that under a combination of deleterious yet realistic conditions, the critical pressure at which fracture takes place can be as low as 15 MPa. These results bring new mechanistic insight into the viability of using the existing natural gas pipeline network to transport hydrogen, and the computational framework presented enables mapping the conditions under which this can be achieved safely.  \\
\end{abstract}

\end{frontmatter}

\date{\today}


\section{Introduction}
\label{sec:intro}
Driven by the prominent role that hydrogen will play as a clean energy vector, there is growing interest in assessing the suitability of transporting gaseous hydrogen through the existing natural gas pipeline network \cite{lipiainenUseExistingGas2023,thawaniAssessingPressureLosses2023,laureysUseExistingSteel2022,galyasEffectHydrogenBlending2023}. From a techno-economic point of view, the desire is to be able to safely and reliably move large volumes of hydrogen at the required gas purity and pressure \cite{penevEconomicAnalysisHighpressure2019a,rongTechnoeconomicAnalysisHydrogen2023}. However, hydrogen is known to dramatically degrade metallic materials \cite{Gangloff2003,Gangloff2012,Djukic2019}. Recent experimental studies have shown that exposing pipeline steels to hydrogen gas (or natural gas/hydrogen mixtures) leads to reduced fracture resistance \cite{san2012technical,boukorttHydrogenEmbrittlementEffect2018b,briottetInfluenceHydrogenOxygen2018,ronevichEFFECTSTESTINGRATE2023}, accelerated fatigue crack growth rates \cite{ronevichHydrogenAcceleratedFatigue2017,ronevichFatigueCrackGrowth2020}, and degraded tensile behaviour (e.g., reductions in breaking stress, notch tensile strength, ductility) \cite{songNotchedtensilePropertiesHighpressure2017,mengHydrogenEffectsX802017}. Consequently, quantifying the factors affecting performance within a robust engineering framework will allow identifying the conditions under which hydrogen can be safely transported in existing and new pipeline infrastructure. This capability would have value for assessing the suitability of existing gas pipelines and provide benefits to new systems. Real-scale trials are ongoing, where pipeline integrity is assessed under a progressively increasing gas pressure and/or hydrogen gas purity \cite{ishikawaIntegrityAssessmentLinepipes2022,kappesBlendingHydrogenExisting2023}; however, these tests are inherently of a relatively short duration and arguably not a good representation of long-term conditions due to a number of reasons, ranging from fatigue damage to the need to destabilise the existing oxide film to enable hydrogen ingress into the pipeline steel. Moreover, there are many different steel grades used in line pipe applications. These are typically fabricated to API 5L Line Pipe specifications and are classified according to tensile properties. Thus, it is impossible to assess all the combinations of steels, welds and operating conditions using physical tests. The goal of the present work is to utilise state-of-the-art computational technologies to overcome these methodological challenges and provide a mechanistic assessment of pipeline integrity as a function of hydrogen pressure and content.\\

Coupled finite element models provide an avenue for predicting hydrogen diffusion and its interplay with material deformation and damage \cite{RILEM2021,Cheng2023}. Typically, these models assume a given hydrogen concentration at the exposed boundaries (although some models can also quantify hydrogen uptake \cite{CS2020b,Hageman2022}), and predict the distribution of hydrogen due to diffusion, driven by gradients of concentration and of hydrostatic stress \cite{Sofronis1989,IJHE2016}, and its sequestration at microstructural traps \cite{Dadfarnia2011,AM2020,IJP2021}. Damage is often assumed to nucleate when a critical combination of mechanical fields and hydrogen content is attained \cite{CupertinoMalheiros2022,Valverde-Gonzalez2022}, and multiple computational technologies have been used to simulate the growth of hydrogen-assisted cracks, including cohesive zone elements \cite{Serebrinsky2004,Moriconi2014,EFM2017,Elmukashfi2020,Colombo2020}, nodal release techniques \cite{Dadfarnia2014}, damage mechanics models \cite{Anand2019,Yu2019b,Depraetere2021}, and phase field methods \cite{CMAME2018,Duda2018,Huang2020,Wu2020b,Golahmar2022,Dinachandra2022}. Recent years have seen significant progress in modelling abilities and it is now possible to deliver hydrogen embrittlement predictions for time and length scales of technological importance \cite{TAFM2020c}, effectively launching a \emph{Virtual Testing} paradigm in hydrogen-containing environments. However, welded joints are often the weakest link in structural components and the aforementioned computational capabilities have yet to be applied to welds. Existing modelling schemes need to be extended to account for the particularities of welds (e.g., residual stresses, microstructural variations, existing defects) to deliver accurate estimations of pipeline integrity in hydrogen-containing environments.\\

In this work, we present a computational framework that couples weld process simulations with deformation-diffusion-fracture modelling of hydrogen-assisted cracking. The approach developed seeks to minimize the number of assumptions required and thus the integrated models perform calculations based on the engineering processes applied during fabrication and exposure. Our focus is on pipeline steels containing seam welds manufactured using submerged arc welding, so as to map safe regimes of operation in hydrogen transport pipelines. Critical pressures are determined for a wide range of scenarios to understand and quantify the role of key variables; namely: residual stresses, microstructure, weld geometry and quality, porosity, existing defects (internal and external), and hydrogen gas pressure. We find that, in the worst (realistic) scenario, H$_2$ pressures as low as 15 MPa can compromise the structural integrity of pipelines.\\

The remainder of this manuscript is structured as follows. In \cref{sec:formulation}, the computational framework is presented, including the weld process modelling stage and the subsequent deformation-diffusion-fracture simulations, as well as their couplings. Then, the results are presented in \cref{sec:results}. Residual stress states, hydrogen distributions and cracking patterns are obtained for two well-characterised seam weld configurations under varying conditions. The quantitative nature of these predictions is strengthened by the ability of the model to predict crack growth resistance behaviour (R-curves) in pipeline steels (X52 and X80), showing a very good match with experiments (see \ref{sec:Rcurve}). The manuscript ends with concluding remarks in \cref{sec:conclusions}.

\section{Modelling framework}
\label{sec:formulation}
\begin{figure}[H] 
    \centering
    \includegraphics[width=\textwidth]{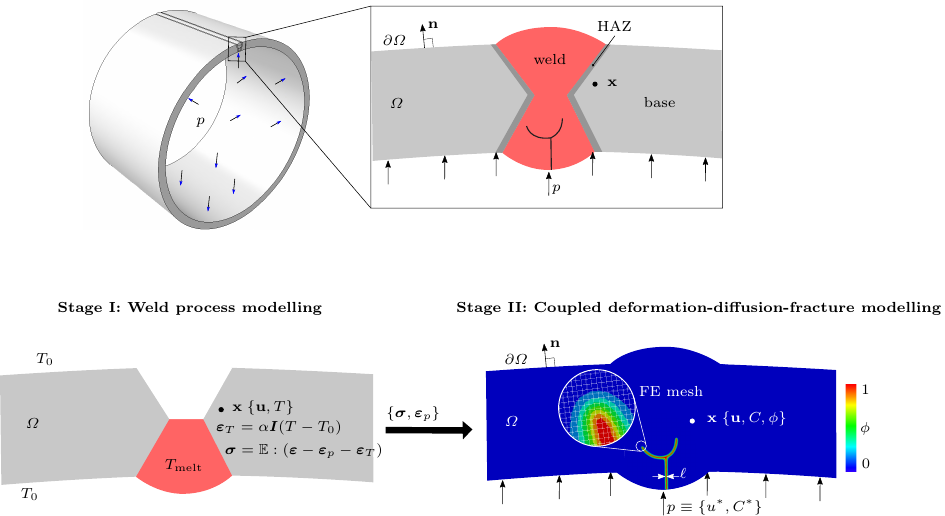}
    \caption{Schematic representation of the finite element modelling framework, aimed at predicting hydrogen-assisted failures of welded joints in pipelines subjected to an internal hydrogen pressure $p$. The first stage of the modelling framework handles the simulation of the welding process, as illustrated here with one weld bead at initial melting temperature $T_\mathrm{melt}$. The second stage uses a coupled, multi-physics phase field formulation to represent the nucleation and growth of cracks, as assisted by hydrogen.}
    \label{fig:methodology}
\end{figure}

As sketched in \cref{fig:methodology}, we present a framework that combines weld process modelling and coupled deformation-diffusion-fracture simulations. The focus is on pipelines containing seam welds manufactured using the submerged arc process and exposed to a hydrogen gas pressure $p$. The weld domain can be divided into three distinct regions: the base metal, the weld metal and the heat affect zone (HAZ). Using established convention, the weld metal is considered the region where complete melting and solidification have taken place, the HAZ describes the extent of the base metal where the heat from welding modifies the original microstructure, and the base metal is the region being unaffected by the welding process carried out. As depicted in \cref{fig:methodology}, the first stage (weld process modelling) is needed to determine the intial stress/strain state of the weld (residual stresses, plastic strain distribution). Then, stage II simulations (embrittlement) can be conducted and cracking can be predicted as a function of the hydrogen gas pressure and relevant weld characteristics. In the following, we describe the theory underlying the weld process modelling (\cref{sec:weld-model}) and the coupled deformation-diffusion-fracture simulations (\cref{sec:HAC-model}), as well as provide details of the numerical implementation (\cref{sec:fem}). 

\subsection{Weld model}
\label{sec:weld-model}
We idealise the welding process as a thermo-mechanical problem. On the weld domain $\Omega$, the independent variables necessary for the description of the model at each point $\mathbf{x}$ and time $t$ are the temperature $T(\mathbf{x},t)$ and the displacement field describing the mechanical deformation of the solid $\mathbf{u}(\mathbf{x},t)$.
The thermal sub-problem is characterised by the heat equation, which reads,
\begin{align}\label{eq:thermal-prob}
    \rho c \pd{T}{t} + \nabla\cdot \left(-k\nabla T \right) = 0 \quad \text{in } \varOmega, \quad 
    {T = T^\ast  \;\;\text{on } \partial\varOmega_T , \quad
    -k\nabla T \cdot \mathbf{n} = q_c + q_r \;\;\text{on } \partial\varOmega_q}, 
\end{align}
where $\rho$, $c$, and $k$ are the mass density, specific heat, and heat conductivity of the material, respectively. 
{We consider convective and radiative cooling of the weldment from all external surfaces $\partial\varOmega_q$ which includes the inner and outer surface of the pipe as well as the weld-base interfaces exposed to the air. The convective cooling is defined using Newton's law $q_c = h_c(T-T_0)$, with room temperature $T_0$ and heat transfer coefficient $h_c$. The radiative heat loss from all external surfaces is modelled using $q_r = \varepsilon_0\sigma_0\big((T-T_\mathrm{abs})^4 - (T_0-T_\mathrm{abs})^4\big)$ with emissivity $\varepsilon_0$, Stefan-Boltzmann constant $\sigma_0$ and absolute temperature $T_\mathrm{abs}$. Moreover, we consider the initial temperature of the weld beads to be the melting point of the material \ie $T(\mathbf{x}, 0) = T_\mathrm{melt}$ and that of the base metal to be the room temperature $T(\mathbf{x}, 0) = T_0$. Considered values of these model parameters are listed in \cref{tab:properties-thermal}. Furthermore, the welding process involves the addition of welding beads in stages along with introducing suitable temperature boundary conditions $T = T^\ast(t)$ at the weld cavity $\partial\varOmega_T$, these specific details are discussed in \cref{sec:ideal-weld-residual-stress}.}

    \begin{table*}[htp]
    \centering
    \setlength\fboxsep{0pt}
    \smallskip%
    \renewcommand\arraystretch{1}
    \begin{tabularx}{0.8\textwidth}{lccccc}
    \toprule
    $h_c$ [Wm$^{-2}$K$^{-1}$] & $\sigma_0$ [Wm$^{-2}$K$^{-4}$] & $\varepsilon_0$ [-] & $T_0$ [$^\circ$C] & $T_\mathrm{melt}$ [$^\circ$C] & $T_\mathrm{abs}$ [$^\circ$C] \\
    \midrule 
    25 & $5.69\times 10^{-8}$ & 0.8 & 20  & 1500 & -273\\
    \bottomrule
    \end{tabularx}
    \caption{{Model parameters for thermal problem.}}
    \label{tab:properties-thermal}
    \end{table*}

Assuming a small strain formulation, the local deformation of the solid is characterised by infinitesimal strain,
\begin{equation}
    \bfepsilon:= \nabla^\mathrm{sym}\mathbf{u} = \frac{1}{2}\Big(\nabla\mathbf{u} + (\nabla\mathbf{u})^T\Big)
\end{equation}
which is additively decomposed into thermal ($\bfepsilon_T$), elastic ($\bfepsilon_e$) and plastic ($\bfepsilon_p$) components; $\bfepsilon = \bfepsilon_e + \bfepsilon_p + \bfepsilon_T$. The thermal strain $\bfepsilon_T$ is computed using the temperature distribution obtained from the thermal sub-problem,
\begin{equation}\label{eq:thermal-strain}
\bfepsilon_T(\mathbf{x},t) := \alpha\boldsymbol{I}(T(\mathbf{x},t) - T_0(\mathbf{x})),
\end{equation}
where $\alpha$ is the coefficient of thermal expansion, $T_0$ denotes the room temperature, and $\boldsymbol{I}$ is the second-order unit tensor.
The kinematics of plasticity are defined by the accumulated plastic strain $\varepsilon_p(\mathbf{x},t):= \int_t | \dot{\bfepsilon}_p(\mathbf{x},t')|\mathrm{d}t'$. Von-Mises plasticity with power law hardening is considered and $\varepsilon_p$ is computed using a conventional return mapping procedure \citep{simo1998-computational}. The elastic strain generates the Cauchy stress $\bfsigma$ which satisfies the global momentum balance. Thus, the mechanical sub-problem of the elastoplastic solid is written as,
\begin{align}\label{eq:mech-prob}
\begin{split}
& \nabla \cdot \bfsigma = \boldsymbol{0} \quad \text{in } \varOmega, \quad
\bfsigma\cdot\mathbf{n} = \boldsymbol{0} \quad \text{on } \partial\varOmega,\\
& \bfsigma = \mathbb{E} : \left(\bfepsilon - \bfepsilon_p - \bfepsilon_T \right), \quad 
||\bfsigma_\mathrm{dev}|| = \sigma_y \left(1 + \dfrac{E\varepsilon_p}{\sigma_y} \right)^{n} \;\;\mathrm{if}\;\dot{\varepsilon}_p > 0,
\end{split}
\end{align}
with deviatoric stress $\bfsigma_\mathrm{dev} := \left(\bfsigma - \frac{1}{3}\trace{\bfsigma}\boldsymbol{I} \right)$, fourth-order elasticity tensor $\mathbb{E}$, Young's modulus $E$, yield strength $\sigma_y$, and hardening exponent $n$.\\  

\begin{figure}[H] 
    \includegraphics[width=\textwidth]{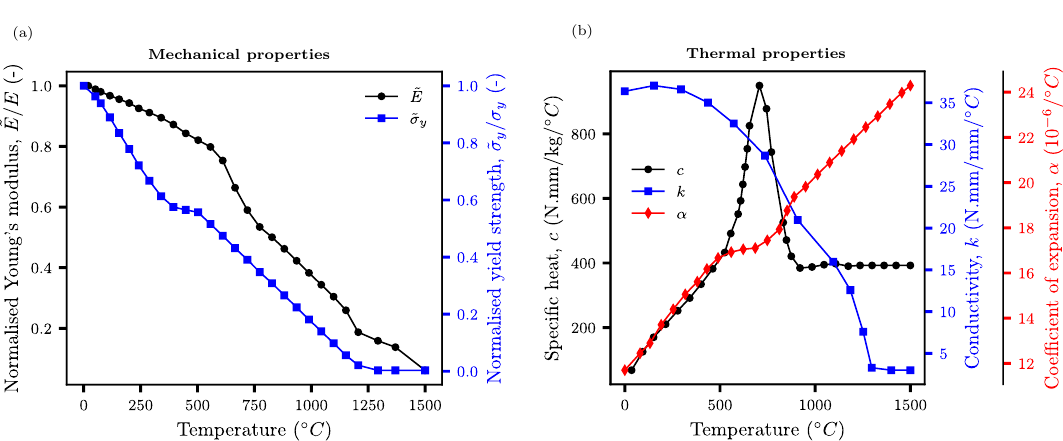}
    \caption{Temperature-dependence of the material properties of carbon steel SA-516 Grade 70, as per ASME, Section II, Part D \citep{ASME_BPVC_IID}. The normalised mechanical properties are presented in (a), where $E$ and $\sigma_y$ denote Young's modulus and yield strength at room temperature, respectively, and $\tilde{\Box}$ represents the corresponding temperature-dependent parameter. The temperature dependence of the thermal properties (specific heat $c$, conductivity $k$, coefficient of thermal expansion $\alpha$) is given in (b).}
    \label{fig:props-tempr}
\end{figure}

{We assume the considered thermo-mechanical problem is weakly coupled where the temperature changes can cause volumetric deformation whereas the thermal processes are unaffected by the deformation of the solid. Therefore}, we have considered a sequential thermo-mechanical procedure with two field variables, temperature $T$ and displacement $\mathbf{u}$. Accordingly, the temperature profile obtained by solving the heat equation, \cref{eq:thermal-prob}, is transferred to the mechanical sub-problem, \cref{eq:mech-prob}, as an initial field, so as to calculate the thermal strain, \cref{eq:thermal-strain}. The temperature dependence of the thermal and mechanical material properties are given in \cref{fig:props-tempr}, following the ASME standard (Section II, Part D) \citep{ASME_BPVC_IID}. The data presented show a monotonic reduction of elastic modulus and yield strength with temperature (\cref{fig:props-tempr}a). In addition, see \cref{fig:props-tempr}b, the thermal conductivity reduces with increasing temperature, whereas the opposite is true for the coefficient of thermal expansion. The specific heat initially increases with temperature up to $700^\circ$C and then decreases rapidly to reach an almost constant value. 

\subsection{Coupled deformation-diffusion-fracture model}
\label{sec:HAC-model}
We idealise the hydrogen-assisted failure of the welded joint as a coupled deformation-diffusion-fracture problem, where the evolving fracture surface is modelled using the phase field method. The variational framework for phase field and its connection with the thermodynamics of fracture are given in \cref{sec:variational}. The particularisation of the model to elastic-plastic fracture is then provided in \cref{sec:elastoplastic-frac}, and its subsequent coupling with hydrogen transport is described in \cref{sec:hydrogen-diffusion}. Finally, the model is enhanced with a phenomenological degradation law for pipeline steels, which is provided in \cref{sec:Gc-hydrogen}.

\subsubsection{Variational framework}\label{sec:variational}
The phase field fracture model is based on Griffith's thermodynamics framework \cite{grifith1920phenomena}, which states that crack growth will occur when a critical strain energy release rate is attained. Considering a deformable solid with strain energy $\Psi(\bfepsilon)$, the variation of the total potential energy $\mathscr{E}$ due to an incremental increase in the crack area $\mathrm{d}A$ is given by,
\begin{equation}\label{eq:frac1}
\vd{\mathscr{E}}{A} = \vd{\Psi (\bfepsilon)}{A} + \vd{W_c (C)}{A}  = 0,
\end{equation}
where the strain tensor $\bfepsilon$ characterises the local deformation and $W_c$ is the work required to create new surfaces. The term $\dd W_c / \dd A$ is also known as the critical energy release rate $G_c$, a material property that characterizes the fracture resistance. This fracture energy is known to be dependent, and very sensitive, to the hydrogen concentration $C$ \cite{Alvaro2015}. Accordingly, Griffith's energy balance can be formulated in a variational form as follows \cite{Francfort1998}
\begin{align}\label{eq:frac2}
   \mathscr{E} = \int_\varOmega \psi (\bfepsilon)\dd\varOmega + \int_\Gamma G_c (C) \dd\Gamma,
\end{align}
with strain energy density $\psi$ and crack surface $\Gamma$. Cracking phenomena can then be predicted as an exchange of stored and fracture energies by minimising \cref{eq:frac2}. However, this requires tracking the crack surface $\Gamma$, which is computationally very challenging \citep{Rabczuk:IAM2013a}. The phase field paradigm provides the necessary framework to predict any arbitrary evolving fracture surface by regularising it using an order parameter $\phi$, which provides a smooth interface between the damaged solid ($\phi = 1$) and intact solid ($\phi = 0$). Upon a suitable choice for the crack surface density function $\gamma$, the Griffith functional (\cref{eq:frac2}) can be approximated by means of the following regularised functional \citep{Bourdin2000,Miehe-1},
\begin{align}\label{eq:frac3}
\begin{split}
  \mathscr{E}_\ell &= \int_\varOmega g (\phi)\psi_0(\bfepsilon)\dd\varOmega + \int_\varOmega G_c (C) \gamma(\phi,\ell) \dd\varOmega, \quad
  \gamma(\phi,\ell) = \left(\dfrac{\phi^2}{2\ell} + \dfrac{\ell}{2} |\nabla\phi|^2\right).  
\end{split}
\end{align}
Here, $\ell$ is a length scale parameter that governs the size of the fracture process zone, $\psi_0$ denotes the strain energy density of the undamaged solid, and $ g (\phi)$ is a degradation function. It can be shown through Gamma-convergence that $\mathscr{E}_\ell$ converges to $\mathscr{E}$ when $\ell\rightarrow 0$ \citep{Chambolle2004}.

\subsubsection{Elastoplastic fracture}
\label{sec:elastoplastic-frac}
Similar to our weld model (\cref{sec:weld-model}), we consider a small strain formulation where the local deformation is characterised by the infinitesimal strain $\bfepsilon:=\nabla^\mathrm{sym}\mathbf{u}$, and it is additively decomposed into the elastic ($\bfepsilon_e$) and plastic ($\bfepsilon_p$) components, as $\bfepsilon = \bfepsilon_e + \bfepsilon_p$. We define the strain energy density of the elastoplastic solid as,
\begin{equation}\label{eq:psi-mech}
\begin{split}
 &\psi :=  g (\phi) \psi_e^{+}\left( \bfepsilon_e \right) + \psi_e^{-}\left( \bfepsilon_e \right) 
+ \bar{ g }(\phi)\psi_p(\varepsilon_p),\\
& \psi_e^{+} = \dfrac{\mathcal{K}}{2}\langle\trace{\bfepsilon_e}\rangle^2 + \mathcal{G}\bfepsilon_e^\mathrm{dev}:\bfepsilon_e^\mathrm{dev}, \quad 
\psi_e^{-} = \dfrac{\mathcal{K}}{2}\langle-\trace{\bfepsilon_e}\rangle^2, \quad
\psi_p = \dfrac{\sigma_y^2}{E(n+1)}\left(1 + \dfrac{E\varepsilon_p}{\sigma_y} \right)^{(n+1)},
\end{split}
\end{equation}
where $\mathcal{K}$ is the bulk modulus, $\mathcal{G}$ denotes the shear modulus, and the strain energy density is divided into its elastic ($\psi_e$) and plastic ($\psi_p$) parts, which are degraded using distinct degradation functions: $ g (\phi)$ and $\bar{ g }(\phi)$, respectively. Quadratic degradation functions are employed, such that
\begin{equation}\label{eq:beta}
 g (\phi) = (1-\phi)^2,\quad
\bar{ g }(\phi) := \beta g (\phi) + (1-\beta),\quad
0\le\beta \le 1.
\end{equation}
where $\beta$ is a parameter that quantifies the amount of plastic work that is stored in the material. Following the seminal work by Taylor and Quinney \cite{taylorLatentEnergyRemaining1933}, we adopt a value of $\beta = 0.1$, as only $10\%$ of the plastic work does not dissipate as heat and is therefore available to create new fracture surfaces. {Hence, the model remains consistent with the thermodynamics of fracture while accounting for the role of plasticity in dissipation and in driving crack growth. Also, the definition of $\bar{ g }(\phi)$ adopted ensures variational consistency.} In addition, we adopt the so-called volumetric-deviatoric split \cite{Amor2009} to decompose the elastic strain energy so as to prevent cracking from nucleating in regions of compressive stress states. As in the welding process modelling stage, we assume elastic-plastic behaviour described by Von Mises plasticity with power law hardening. Considering also an isotropic degradation of the stress \cite{Ambati2016}, the variation of the energy functional $\mathscr{E}_\ell$ with respect to the displacement $\delta\mathbf{u}$ renders
\begin{align}\label{eq:eqn-disp}
\begin{split}
& \nabla \cdot \bfsigma = \boldsymbol{0}, \qquad \bfsigma =  g (\phi)\mathbb{E} : \left(\bfepsilon - \bfepsilon_p \right), \\
& ||\bfsigma_\mathrm{dev}|| = \bar{ g }(\phi)\sigma_f(\varepsilon_p)\;\;\mathrm{if}\;\dot{\varepsilon}_p > 0,\quad 
\sigma_f(\varepsilon_p) := \sigma_y \left(1 + \dfrac{E\varepsilon_p}{\sigma_y} \right)^{n}. 
\end{split}
\end{align}
While by taking the variation of the energy functional $\mathscr{E}_\ell$ with respect to the phase field $\delta\phi$ one reaches
\begin{align}\label{eq:eqn-phi}
\begin{split}
& \nabla^2 \phi = \left( \dfrac{\phi}{\ell^2} +  g '(\phi) \dfrac{\mathcal{H}}{\ell G_c}  \right), \qquad
\mathcal{H} := \max{\left( \max{\left(\psi_{e}^{+}, \psi_{e(t)}^{+} \right)} + \beta\psi_p \right)},
\end{split}
\end{align}
where damage irreversibility has been enforced by adopting a non-decreasing crack driving energy $\mathcal{H}$ \citep{Miehe-2}. It is also worth noting that the phase field length scale parameter is directly related to the fracture strength $\hat{\sigma}$ as \cite{PTRSA2021},
\begin{equation}\label{eq:at2-l0}
    \ell = \dfrac{27}{256}\dfrac{EG_c}{\hat{\sigma}^2}.
\end{equation}

\subsubsection{Hydrogen transport}
\label{sec:hydrogen-diffusion}
We describe the diffusion of hydrogen through the crystal lattice by employing the usual Fickian model for the hydrogen flux $\mathbf{J}$, such that
\begin{equation}
    \pd{C}{t} = -\nabla\cdot \mathbf{J}, \quad \text{with} \quad
    \mathbf{J} = - \dfrac{DC}{RT} \nabla\mu,
\end{equation}
with (apparent) hydrogen diffusion coefficient $D$, ideal gas constant $R = 8.314$ J/(mol K), and absolute temperature $T$. 
We define the chemical potential $\mu$ assuming that the hydrogen resides in interstitial lattice sites of steel as a diluted species,
\begin{align}\label{eq:psi-chem}
\mu = \mu^0 + RT\ln{\left(\dfrac{\theta_L}{1-\theta_L}\right)} - \bar{V}_H \sigma_h,
\end{align}
where $\mu^0$ is the reference chemical potential, $\sigma_h:=\trace{\bfsigma}/3$ is the hydrostatic stress, and $\bar{V}_H$ is the partial molar volume of hydrogen in solid solution, which takes a value of $2000\;\mathrm{mm}^3/\mathrm{mol}$ in iron-based materials. Also, $\theta_L=C/N_L$ is the lattice occupancy, a function of the lattice hydrogen concentration and the lattice site density $N_L$. {The role of traps is here accounted for by taking $D$ to be the effective diffusion coefficient, but the model can readily be extended to explicitly consider multiple trap types (see Ref. \cite{IJP2021})}.\\

It is also necessary to handle the {interplay between hydrogen transport and crack growth.} It is expected that the hydrogen gas will {very promptly} occupy the space created by crack advance, {exposing the newly created crack surfaces to the environmental hydrogen concentration}. Accordingly, the diffusivity of material points in the damaged region is enhanced by making use of an amplification coefficient $k_d \gg 1$ and a damage threshold, $\phi_\mathrm{th}$, above which the hydrogen environment is assumed to expand. Then, the hydrogen diffusion coefficient can be expressed as,
\begin{align}\label{eq:artificial-diffusivity}
    D = D_0 \big( 1 + k_d\langle \phi - \phi_\mathrm{th} \rangle \big)
\end{align}
where $D_0$ is the hydrogen diffusivity of the undamaged material and $ \langle \Box \rangle$ are Macaulay brackets. {In this regard, it should be noted that $\phi_\mathrm{th}$ denotes the damage threshold above which an interconnected network of cracks is assumed to exist, such that the hydrogen can find its way from the environment. Exploiting the analogy between phase field and damage mechanics, we assume $\phi_\mathrm{th} \approx 0.8$ to be a sensible choice (microcracks occupy 80\% of the surface plane). On the other side, $k_d$ is a numerical parameter that should be taken to be as large as possible. We find that $k_d \approx 10^4$ delivers a robust model and is sufficient to achieve the desired effect.} The reader is referred to Ref. \citep{andreasDiaz-comsol-pfm} for further details of this approach and its numerical implementation, and to Ref. \cite{Martinez-Paneda2020} for a discussion on alternative approaches such as penalty-based \emph{moving} chemical boundary conditions.

\subsubsection{Hydrogen degradation function}
\label{sec:Gc-hydrogen}
\begin{figure}[!htb]
    \centering
    \includegraphics[width=0.9\textwidth]{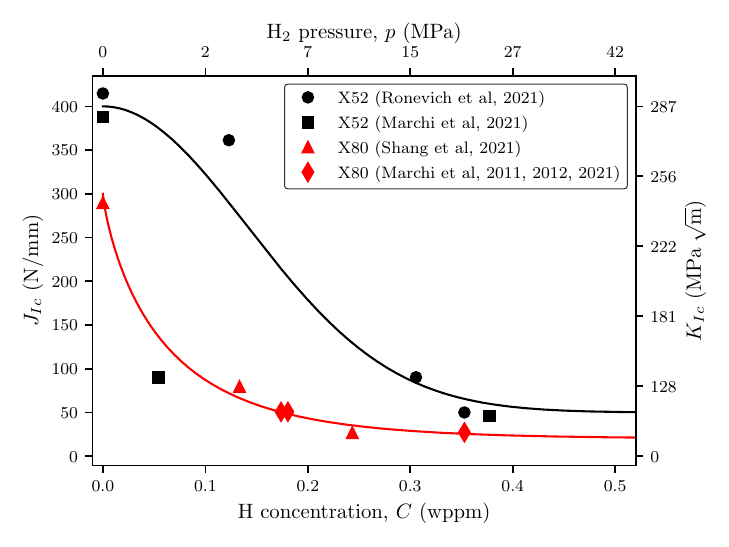}
    \caption{Hydrogen-dependent critical fracture energy for API steels: X52 \citep{ronevichMaterialsCompatibilityConcerns2021,sanmarchiMaterialsEvaluationHydrogen2021}, X80 \citep{shangDifferentEffectsPure2021,sanmarchiFractureFatigueCommercial2011,sanmarchiFractureResistanceFatigue2012,sanmarchiHYDROGENCOMPATIBILITYSTRUCTURAL2021}. The fitted curves are obtained using \cref{eq:hydrogen-Gc} with fitting parameters listed in \cref{tab:database}. 
    For convenience, the corresponding critical stress intensity factor $K_{Ic}$ is added on the right and it is related to $J_{Ic}$, as $K_{Ic} = \sqrt{E'J_{Ic}}$ with $E' = E/(1-\nu^2)$ for plane strain conditions.
    Also the corresponding hydrogen pressure $p$ is added on top; lattice hydrogen concentration $C$ and hydrogen pressure $p$ are related via Sievert's law $C = S\sqrt{p}$, with the solubility of steel taken to be $S = 0.077\;\mathrm{wppm}\;\mathrm{MPa}^{-0.5}$ \citep{martinHydrogenEmbrittlementFerritic2020}.}
    \label{fig:database}
    \end{figure}

It remains to define the dependency of the fracture energy upon the hydrogen content. To this end, a hydrogen degradation function $f(C)$ has to be defined, such that,
\begin{equation}
    G_c (C) = f (C) G_c (0)
\end{equation}
where $G_c (0)$ is the critical energy release in a hydrogen-free environment. Several approaches can be adopted here, including mechanistic ones (e.g., using atomistic predictions of surface energy reduction with hydrogen coverage \cite{CMAME2018}) and phenomenological ones. Since the relationship between fracture energy and hydrogen content is well characterised in pipeline steels, we choose to adopt a phenomenological approach. To this end, an analogy can be established between the $J$ integral and the energy release rate $G$ under the assumption of small scale yielding conditions, with $G_c$ being the value of $J$ at which crack growth initiates (see \ref{sec:Rcurve} for details). Experimental data exists characterising the drop of fracture energy with hydrogen content for pipeline steels - \cref{fig:database} shows a collection of published data for X52 and X80 pipeline steels. The data is provided for the critical value of $J$ under mode I fracture conditions ($J_{Ic}$), which is measured for a crack extension of 0.2 mm. 
As shown by the curves in \cref{fig:database}, the data can be fitted through the following exponential function,
\begin{equation}
\label{eq:hydrogen-Gc}
J_{Ic}(C) = f(C) J_{Ic} (0)= \left[ \frac{J_{Ic}^{\text{min}}}{J_{Ic} (0)}  + \left( 1 - \frac{J_{Ic}^{\text{min}}}{J_{Ic} (0)} \right) \exp{ \left( - q_1 C^{q_2} \right)} \right] J_{Ic} (0) 
\end{equation}
where $q_1$ and $q_2$ are fitting parameters, and $J_{Ic}^{\text{min}}$ is the saturation magnitude, the lowest value of $J_{Ic}$ in hydrogen-containing environments. The values of $q_1$ and $q_2$ that provided the best fit to the experimental data are given in \cref{tab:database}, together with the measured values of $J_{Ic}$, $J_{Ic}^{\text{min}}$, and yield strength $\sigma_y$. As shown in \ref{sec:Rcurve}, these choices are found to accurately reproduce the crack growth resistance of pipeline steels in the presence of hydrogen. \ref{sec:Rcurve} also discusses the role of fracture strength $\hat{\sigma}$ in influencing the degree of plastic dissipation with crack growth. {It should be noted that the fracture energy versus hydrogen content data available for pipeline steels refers mostly to the base metal. Some data exists for weld metal suggesting a similar behaviour \cite{ronevichHydrogenassistedFractureResistance2021} and thus the same degradation law, Eq. (\ref{eq:hydrogen-Gc}), is here adopted to describe the hydrogen degradation of material points in the weld, base and HAZ regions, albeit with different coefficients. Nevertheless, a more detailed experimental quantification would be very helpful in reducing modelling uncertainties.}

    \begin{table*}[htp]
    \centering
    \setlength\fboxsep{0pt}
    \smallskip%
    \renewcommand\arraystretch{1}
    \begin{tabularx}{0.7\textwidth}{lccccc}
    \toprule
    Steel & $\sigma_y$ [MPa] & $J_{Ic}(0)$ [N/mm] & $J_{Ic}^{\text{min}}$ [N/mm] & $q_1$ [-]  & $q_2$ [-]\\
    \midrule 
    X80  & 660 & 289  &  20 & 9 & 0.8 \\
    X52  & 430 & 400  &  50 & 25 & 2 \\
    \bottomrule
    \end{tabularx}
    \caption{Material parameters characterising the behaviour of pipeline steels (X52, X80) in the presence of hydrogen. $J_{Ic}(0)$ denotes the measured value of $J_{Ic}$ in the absence of hydrogen, $J_{Ic}^{\text{min}}$ is the lowest value of $J_{Ic}$ measured for that material in a hydrogen-containing environment, and $q_1$ and $q_2$ are fitting parameters in \cref{eq:hydrogen-Gc}.}
    \label{tab:database}
    \end{table*}

\subsection{Numerical implementation}
\label{sec:fem}
We numerically implement the thermo-mechanical model of the welding process (\cref{sec:weld-model}) and the coupled deformation-diffusion-fracture model for welded joints (\cref{sec:HAC-model}) using the commercial finite element package \texttt{Abaqus}. Several user subroutines were developed. For the welding process modelling, a user material (\texttt{UMAT}) subroutine is used to implement power law temperature-dependent elastoplasticity, extending the code provided in Ref. \cite{EJMAS2019b}. The welding sequence is modelled by removing all the elements in the weld region at the beginning of the simulation and then progressively adding (or activating) the elements corresponding to each weld bead. This is achieved using the `*Model Change' feature in \texttt{Abaqus}. Moreover, the required temperature boundary conditions corresponding to each weld pass are automated using a welding interface plugin \citep{ahmedNumericalSimulationSteel2019}. The residual stresses and strains computed in the weld process modelling stage are provided as input to the coupled deformation-diffusion-fracture simulations using a \texttt{SDVINI} subroutine. Specifically, we transfer the accumulated plastic strain, and the elastic and plastic parts of the strain tensor ($ \varepsilon_p, \bfepsilon_e, \bfepsilon_p$) as the initial values. In regards to the hydrogen-assisted fracture simulations, a \texttt{UMAT} subroutine is developed to couple elastic-plastic constitutive behaviour and phase field fracture. The implementation exploits the analogy between the heat equation and the phase field evolution equation, extending to elastic-plastic solids the work by Navidtehrani et al. \cite{navidtehraniSimpleRobustAbaqus2021,Materials2021}. Hydrogen transport is simulated using a user element (\texttt{UEL}) subroutine, where the gradient of hydrostatic stress is computed by an extrapolation-based method \citep{IJSS2015,pfczm-hydrogen2}. The \texttt{UMAT} and \texttt{UEL} subroutines are coupled using Fortran modules. The codes developed and pre-/post-processing Python scripts with examples are available for download at \url{https://www.empaneda.com/codes/}\footnote{To be uploaded immediately after publication}.

\section{Results} 
\label{sec:results}
\begin{figure}[H] 
    \centering
    \includegraphics[width=\textwidth]{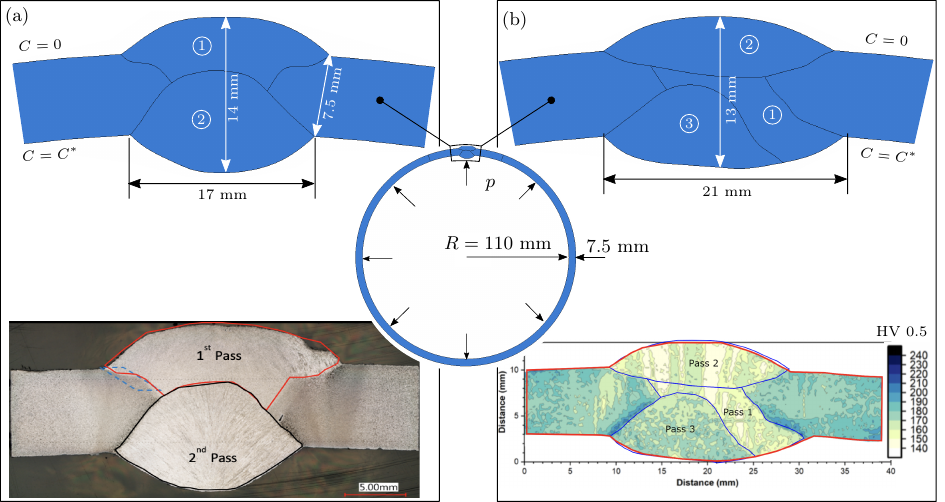}
    \caption{Schematic representation of the boundary value problem considered, a pipeline containing a longitudinal weld. Two weld configurations were employed based on images of ex-service welds from natural gas pipelines: (a) a two-pass weld of dimensions 17 x 14 mm, and (b) a three-pass weld of dimensions 21 x 13 mm. The microstructural heterogeneity of the second weld was characterised using a Vickers Hardness (VH) mapping approach. The associated chemical boundary conditions, including the internal exposure to a hydrogen concentration $C^*$, are also displayed.}
    \label{fig:ideal-weld-geo}
\end{figure}

As shown in \ref{sec:Rcurve}, the coupled deformation-diffusion-fracture model presented in \cref{sec:HAC-model} can accurately capture the crack growth resistance of X52 and X80 pipeline steels in air and hydrogen-containing environments. This validated model is now coupled with welding process simulations and used to gain insight into the structural integrity of seam (i.e., longitudinal) welds in a representative pipeline with an inner radius of 110 mm and thickness of 7.5 mm. As shown in \cref{fig:ideal-weld-geo}, the pipeline contains a seam weld that is symmetrically positioned at the top of the pipeline section. We shall consider two weld configurations, based on images of seam welds from gas transport pipelines taken out of service. The two welds, shown in \cref{fig:ideal-weld-geo}, are representative of submerged arc welded joints typical of pipelines considered for hydrogen transport and span a relevant range of seam weld configurations and geometries. The first weld to be considered is a two-pass seam weld (\cref{Sec:TwoPass}), while the second one is a three-pass weld. Detailed microstructure (hardness) information is available (see \cref{fig:ideal-weld-geo}), allowing us to investigate the influence of spatially-varying properties (\cref{Sec:ThreePass}). These two realistic weld configurations are used to quantify the critical pressures at which hydrogen-assisted fractures compromise the structural integrity of the pipeline and to investigate the role of weld geometry, existing defects, porosity, residual stresses, weld/base metal mismatch, weld process (number of passes), and microstructural heterogeneity. In both cases, simulations were conducted for two pipeline steels: X52 and X80. The material properties employed for each steel are given in \cref{tab:weld1-param}.

    \begin{table*}[htbp]
    \centering
    \setlength\fboxsep{0pt}
    \smallskip%
    \setlength{\tabcolsep}{3.5pt}
    \renewcommand\arraystretch{1}
    \begin{tabularx}{\textwidth}{cccccccccc}
    \toprule
    & & $E$ (MPa) & $\nu$ (-) & $\sigma_y$ (MPa) & $n$ (-) & $G_{c}(0)$ (N/mm) & $G_{c}^\mathrm{min}$ (N/mm) & $\ell$ (mm) & $D$ (mm$^2$/s)\\
    \midrule 
    \multirow{2}{*}{X80} 
    & Base  & \numprint{187000} & 0.3 & 660   &   0.10  & 60 & 7  & 0.17  & $4.5\times 10^{-4}$ \\
    & Weld  & \numprint{196350} & 0.3 & 726   &   0.05  & 54 & 6.3 & 0.13 & $3\times 10^{-4}$ \\
    \midrule
    \multirow{2}{*}{X52} 
    & Base    & \numprint{187000} & 0.3 & 430   &   0.10  & 60 & 16  & 0.40  & $4.5\times 10^{-4}$ \\
    & Weld    & \numprint{196350} & 0.3 & 473   &   0.05  & 54 & 14.4 & 0.31 & $3\times 10^{-4}$ \\
    \bottomrule
    \end{tabularx}
    \caption{Material parameters adopted for the weld and base metal regions of the X52 and X80 pipeline steels used in the analyses.}
    \label{tab:weld1-param}
    \end{table*}
The material properties have been chosen based on experimental measurements from the literature and current understanding of weld region characterisation. The Young's modulus of the weld is usually higher than that of the base metal \citep{arpin2003material}, and accordingly is here taken to be $5\%$ higher than that of the base metal. The weld is usually stronger and comparatively brittle compared to the base metal due to the thermal loading history; therefore we have considered the yield strength of the weld to be 10\% higher than that of the base metal, whereas the fracture energy is 10\% lower. In regards to the work hardening capacity, studies on pipeline steels report a higher magnitude for the base metal relative to the weld material \cite{ronevichHydrogenassistedFractureResistance2021}, and accordingly, this is accounted for by using a lower hardening exponent $n$ for the weld ($n=0.1$ vs $n=0.05$). The diffusivity values of hydrogen in pipeline steel for both weld and base metal are taken from Ref. \cite{souzaEffectMicrostructureHydrogen2017}. The fracture energy and fracture strength parameters, and their sensitivity to hydrogen concentration, are introduced following the experimental data and simulations presented in \cref{sec:Gc-hydrogen} and \ref{sec:Rcurve}. Plane strain conditions are assumed. 


\subsection{Case study 1: two-pass seam weld}
\label{Sec:TwoPass}

We begin our analysis by describing the weld process simulation in the two-pass weld (\cref{sec:ideal-weld-residual-stress}), and then proceed to predict the critical pressure at which hydrogen compromises the structural integrity of the weld, considering the role of residual stresses, weld porosity, and the presence of internal and external defects (\cref{sec:ideal-weld-hac}). 

\subsubsection{Welding residual stress}
\label{sec:ideal-weld-residual-stress}
\begin{figure}[H] 
    \centering
    \includegraphics[width=\textwidth]{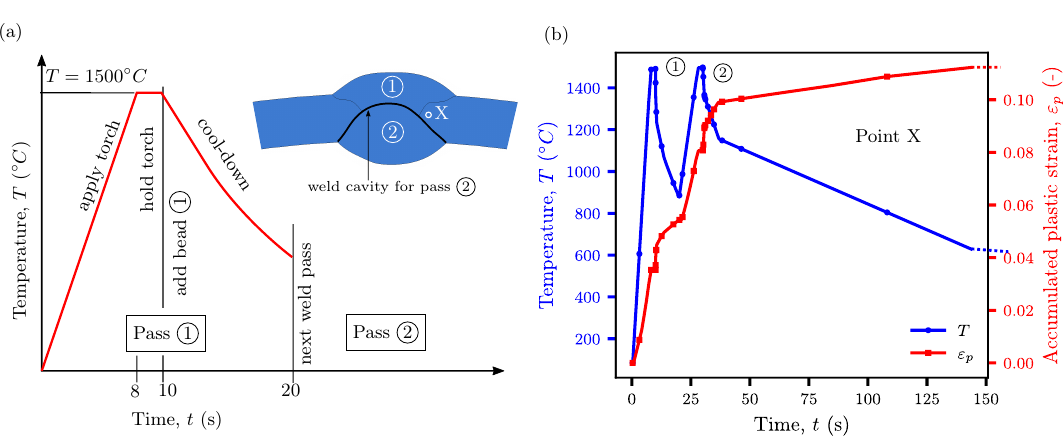}
    \caption{Two-pass weld welding process modelling: (a) schematic representation of the temperature boundary conditions for a two-pass weld; and (b) thermal loading history and corresponding accumulated plastic strain at point `X' near the weld. Only the first 150 seconds are shown but the simulation is continued until reaching room temperature ($20\degree$C).}
    \label{fig:ideal-weld-thermal}
\end{figure}

\begin{figure}[!ht] 
    \centering
    \includegraphics[width=\textwidth]{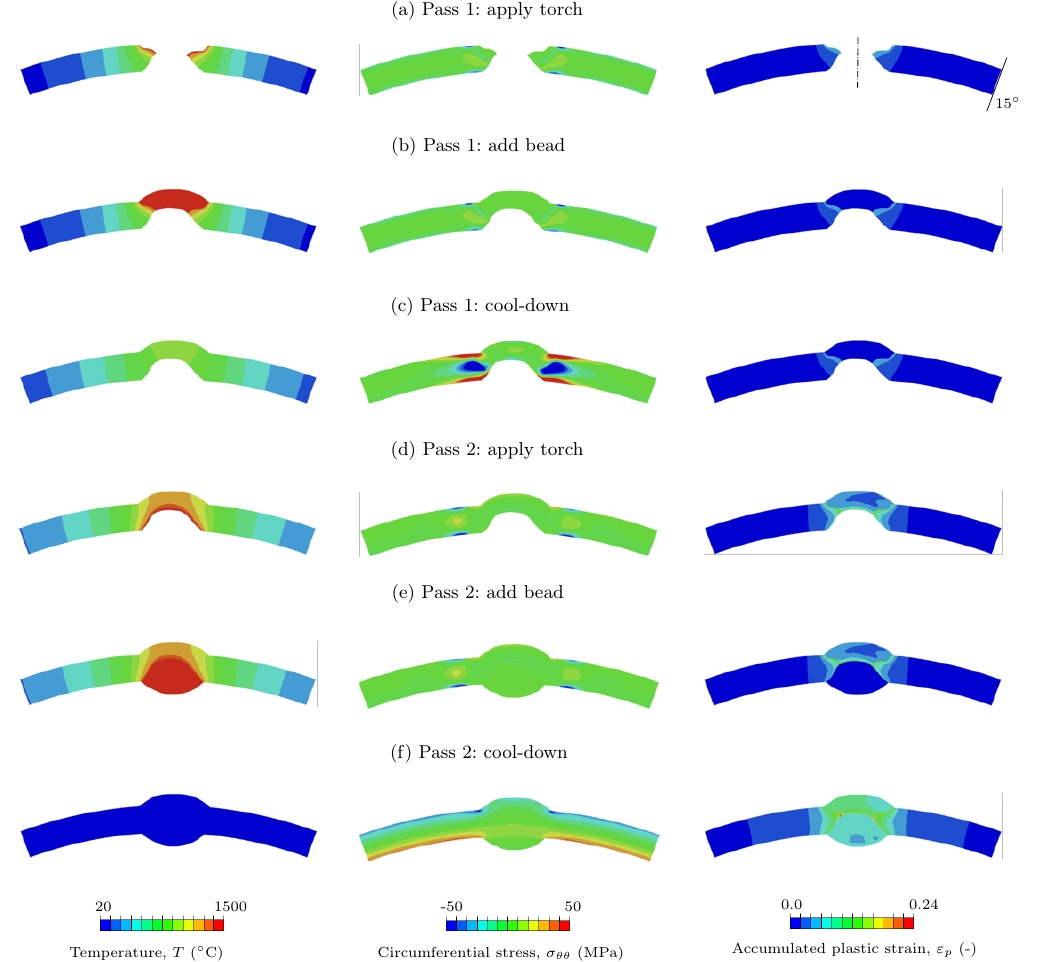}
    \caption{Evolution of temperature, residual stress and plastic strain in the two-pass welding process. This set of representative results has been obtained for X52 pipeline steel.}
    \label{fig:ideal-weld-residual-stress}
\end{figure}

The simulation steps of the two-pass welding process are sketched in \cref{fig:ideal-weld-thermal}a. This involves applying temperature boundary conditions at the weld cavity, followed by adding molten weld bead (at $1500^\circ\mathrm{C}$), and allowing it to cool down for 10 s before applying temperature boundary conditions at the weld cavity for the next weld pass. This is implemented as a sequential thermal-mechanical simulation in \texttt{Abaqus} (see \cref{sec:formulation}). 
A typical material point close to the weld experiences changes in mechanical fields as a result of the thermal load cycles. This is shown in \cref{fig:ideal-weld-thermal}b for a representative material point near the weld. As it can be observed, each weld pass brings a peak in temperature and the associated thermal expansions and contractions generate thermal residual stresses. This, together with the temperature dependence of the mechanical properties (\cref{fig:props-tempr}a), results in material yielding and in a monotonic increase in accumulated plastic strain until the room temperature plateau is reached. The spatial variation of residual stresses and strains at different stages of the welding process is presented in \cref{fig:ideal-weld-residual-stress}.
During the application of the welding torch and the addition of molten weld metal for the first pass (Figs. \ref{fig:ideal-weld-residual-stress}a-b), the temperature increases significantly over a region wider than the weld domain, which results in thermal expansion and corresponding compression in the circumferential direction. This is accompanied by the accumulation of plasticity near the weld cavity (i.e., near the location of the torch) due to the thermal gradient and the corresponding reduction of yield strength with increasing temperature (see Figs. \ref{fig:ideal-weld-residual-stress}a-b). The pipeline section then experiences circumferential tensile stretching at the periphery due to the cooling of the weld bead for the first pass (\cref{fig:ideal-weld-residual-stress}c). The application of the torch at the weld cavity and the addition of molten weld metal for pass 2 reverses the circumferential stress state in the pipeline section (see Figs. \ref{fig:ideal-weld-residual-stress}d-e) and brings as well a significant accumulation of plasticity along the HAZ. The plastic region is further expanded due to the thermal contraction that takes place during the final cooling stage (\cref{fig:ideal-weld-residual-stress}f). This results in tensile and compressive circumferential stresses at, respectively, the inner and outer faces of the pipeline, following the curvature of the pipeline geometry. {The heterogeneous nature of the weld is thus characterised in this first case study through the different properties of base and weld metals, as listed in Table \ref{tab:weld1-param}, and through the initial residual stress/strain distribution resulting from the welding process. This is unlike the second case study, where a hardness map is available that enables assessing the role of microstructural heterogeneities such as the influence of varying properties along the HAZ region.}

\subsubsection{Hydrogen embrittlement predictions}
\label{sec:ideal-weld-hac}
The embrittlement and fracture resistance of the welded pipeline are characterised by subjecting the welded pipeline section to a monotonically increasing hydrogen pressure $p$. This was simulated by applying, as Dirichlet boundary conditions, an incremental hydrogen concentration $C^\ast$ and radial displacement $u^\ast_r$ at the inner face of the pipe. The former is determined by making use of Sievert's law, such that,
\begin{equation}
\label{eq:pressure}
    C^\ast = S\sqrt{p},
\end{equation}
\noindent where the solubility of hydrogen in steel is $S=0.077\; \mathrm{wppm}\;\mathrm{MPa}^{-0.5}$ \citep{martinHydrogenEmbrittlementFerritic2020}. While the radial displacement is defined by exploiting the linear elastic relation between pressure and radial displacement of a thin cylinder \citep{timoshenko1959-plate-shell},
\begin{equation}
\label{eq:pressure-displacement}
    u^\ast_r = \dfrac{pR^2}{bE}\left( 1 - \dfrac{\nu}{2} \right), 
\end{equation}
where $R$ is the inner radius of the pipeline and $b$ is the pipeline thickness. A very slow loading rate of 21 Pa/s was adopted to be on the conservative side and suppress rate effects. At the outer surface, a boundary condition of $C=0$ was adopted, as the outer surface of transport pipelines is typically exposed to environments with negligible hydrogen content. As gas transport pipelines are designed to carry out pressures below the plastic yielding of the material, we consider two potential outcomes: failure due to plastic yielding or failure due to cracking. The yielding pressure can be estimated analytically from the pipeline dimensions as,
\begin{equation}
    p_y = \frac{\sigma_y b}{R}
\end{equation}
which results in yielding pressures of 45 MPa for the X80 pipeline and of 29 MPa for the X52 steel. Despite its intrinsic assumptions, this analytical estimate is not far from the finite element predictions and similar values of yielding pressures are obtained in the numerical calculations if a global yielding criterion such as $\varepsilon_p \geq 0.02$ in the base metal is adopted. Calculations were arrested if cracking had not taken place before the aforementioned critical yielding pressures were reached, ensuring the validity of \cref{eq:pressure-displacement}.\\

\begin{figure}[H] 
    \centering
    \includegraphics[width=\textwidth]{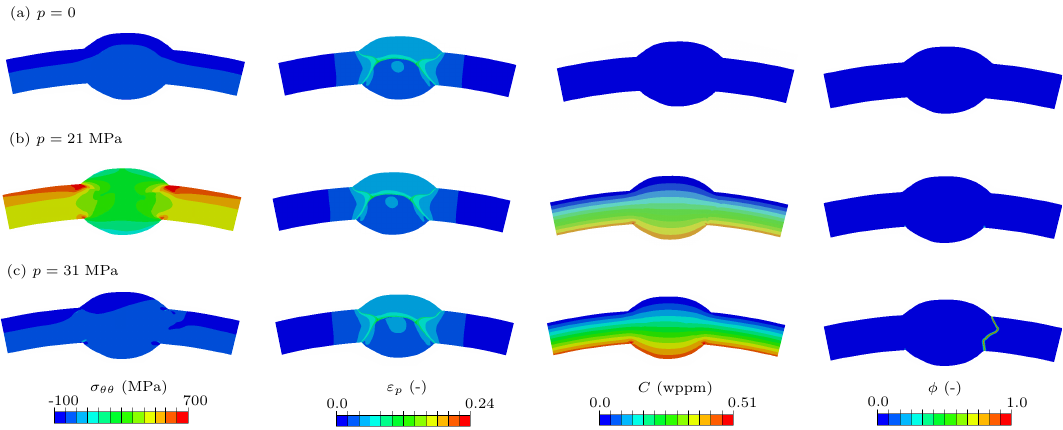}
    \caption{Predictions of circumferential stress ($\sigma_{\theta\theta}$), accumulated plastic strain ($\varepsilon_{p}$), lattice hydrogen concentration ($C$), and phase field fracture ($\phi$) as a function of the pipeline pressure: (a) initial state ($p=0$), (b) $p=21$ MPa, and (c) $p=31$ MPa. Results obtained for the two-pass weld considering X80 pipeline steel as the base and weld materials.}
    \label{fig:ideal-weld-evol-x80}
\end{figure}

The evolution of stress, plasticity, hydrogen concentration, and phase field fracture are shown in \cref{fig:ideal-weld-evol-x80} for the representative case of the X80 welded joint. The initial state (\cref{fig:ideal-weld-evol-x80}a) is hydrogen-free and shows no damage, but welding residual stresses and plastic strains are present. As the applied pressure increases (\cref{fig:ideal-weld-evol-x80}b), we observe an increase in circumferential stresses and hydrogen content. The hydrogen distribution varies monotonically from the inner concentration to the $C = 0$ magnitude of the outer surface. The different diffusivities of weld and base material do not appear to have a significant impact on the hydrogen distributions, given the slow loading rate adopted. In this regard, it is worth noting that the magnitude of the residual tensile hydrostatic stress resulting from the welding process is not high enough to accumulate hydrogen within the heat-affected zone (HAZ) under the conditions considered here. As the pressure keeps rising (\cref{fig:ideal-weld-evol-x80}c), higher stresses and hydrogen concentrations are predicted and this eventually leads to the nucleation of a crack near the weld-base metal junction that goes on to propagate until reaching the outer surface, degrading the load carrying capacity of the pipeline. This crack nucleates at the inner surface of the pipeline section, where the hydrogen content is higher and thus the material toughness is lower. For this weld configuration and material (X80 pipeline steel), cracking takes place at an applied pressure of 31 MPa (i.e., much before yielding). However, X52 is comparatively more ductile and yields before cracking is predicted.

\begin{figure}[H] 
    \centering
    \includegraphics[width=\textwidth]{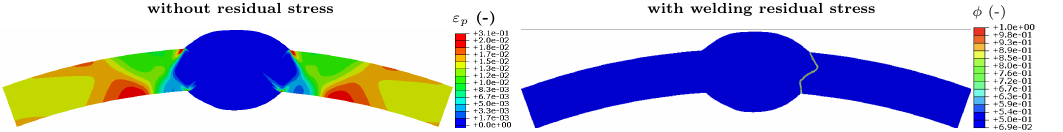}
    \caption{Investigating the role of welding residual stresses on the failure of hydrogen transport pipelines: in the absence of residual stresses plastic failure occurs (left, $\varepsilon_p$ contours), while when residual stresses are accounted for cracking is observed (right, $\phi$ contours). Results obtained for the two-pass weld considering X80 pipeline steel as the base and weld materials.}
    \label{fig:ideal-weld-residual-stress-effect}
\end{figure}

We shall now proceed to investigate the influence of residual stresses, so as to determine the importance of optimising the welding process. \cref{fig:ideal-weld-residual-stress-effect} shows the failure predictions obtained for the X80 pipeline steel case for simulations accounting for and neglecting residual stresses. It can be seen that in the absence of residual stresses, the weld pipeline section fails due to yielding, at $p \approx 45$ MPa, while incorporating residual stresses results in cracking at lower pressures ($p \approx 31$ MPa). Hence, the results suggest that weld quality can be of significant importance and that structural integrity predictions must account for the role of welding residual stresses.\\ 

\begin{figure}[H] 
    \centering
    \includegraphics[width=\textwidth]{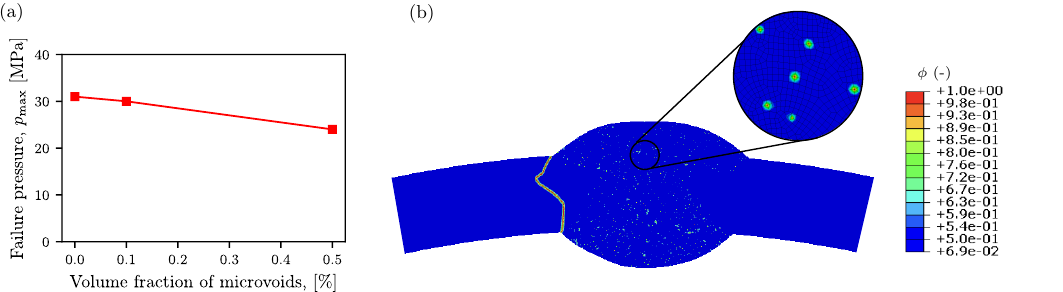}
    \caption{Investigating the role of porosity on the failure of hydrogen transport pipelines: (a) sensitivity of the critical failure pressure to the void volume fraction, and (b) cracking pattern (phase field contours) observed for a volume fraction of 0.5\%, which corresponds to a critical pressure of $p_\mathrm{max}=24$ MPa. Results obtained for the two-pass weld considering X80 pipeline steel as the base and weld materials.}
    \label{fig:ideal-weld-microvoid}
\end{figure}

Another aspect to consider is the potential presence of voids (pores) that can result from the welding process due to gas trapping. Typical diameters for pores resulting from welding gas trapping are within the range of 4 to 10 $\mu$m. Hence, we introduce a random distribution of voids with an approximate size of 7 $\mu$m and vary their number to consider various volume fractions, ranging from 0 to 0.5\%. The voids are introduced by defining an initial condition of $\phi = 1$ on randomly selected nodes. The results obtained are presented in \cref{fig:ideal-weld-microvoid}, showing both the sensitivity of the critical failure pressure to the void volume fraction and the phase field contours, so as to observe cracking patterns and initial void distributions. The results show a change in cracking pattern, with cracks still originating near the inner surface (where the hydrogen content is higher) and next to the weld-base material junction, but with the crack trajectory being influenced by the presence of voids. The critical failure pressure appears to be sensitive to the volume fraction of microvoids, with this trend accelerating for high microvoid volume fractions such that the critical pressure drops from 31 MPa to 24 MPa for a void volume fraction of 0.005. In the case of the X52 pipeline steel case, the presence of voids of 7 $\mu$m diameter with volume fractions of 0.005 or smaller does not change the failure mode; the X52 pipeline fails by yielding independently of the presence of voids. This outcome is similar to the fact that the X52 behaviour was not sensitive to welding residual stress.

\begin{figure}[H] 
    \centering
    \includegraphics[width=\textwidth]{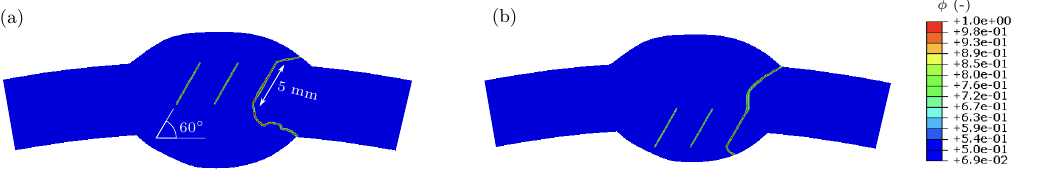}
    \caption{Investigating the role of internal pre-existing defects on the failure of hydrogen transport pipelines. Phase field contours for three internal flaws aligned 60$^\circ$ relative to the horizontal direction under two scenarios: (a) located near the centre of the pipeline section ($p_\mathrm{max} = 23$ MPa), and (b) located near the inner surface ($p_\mathrm{max} = 22$ MPa). Results obtained for the two-pass weld considering X80 pipeline steel as the base and weld materials.}
    \label{fig:ideal-weld-flaw-internal}
\end{figure}

An important concern when assessing the viability of using the existing natural gas pipeline network to transport hydrogen is the possible presence of existing defects and the role that they can play in lowering the admissible pressures. Internal and near-surface cracks of up to 3-5 mm in length are expected to be present in existing infrastructure. Therefore, we conducted simulations for different pre-existing defect scenarios to assess their impact on performance. As with the pores, these defects were introduced by defining as initial condition $\phi=1$ on selected nodes. First, the evaluation was carried out on the X80 pipeline steel case. \cref{fig:ideal-weld-flaw-internal} shows the results obtained in the presence of three 5 mm long internal flaws oriented at $60^\circ$ relative to the horizontal axis. Two scenarios were considered, one where the flaws were located in the centre region of the weld - roughly equidistant to the inner and outer surfaces (\cref{fig:ideal-weld-flaw-internal}a), and another where the flaws were still internal (not exposed to the surface) but much closer to the inner surface (\cref{fig:ideal-weld-flaw-internal}b). In the former scenario, crack growth occurs from both the top and bottom of the flaw closest to the HAZ resulting in a brittle fracture at a hydrogen pressure of $p_{\text{max}}=23$ MPa. Cracking was predicted at a pressure of $p_{\text{max}}=22$ MPa, for the case of the near-surface flaws. The results suggest that the location of the flaw relative to the HAZ is a key factor, more critical than the distance between the flaw and the inner surface, and that the presence of internal defects can reduce the maximum admissible pressure by 20-30\% relative to the defect-free scenario (where $p_{\text{max}}=31$ MPa, see \cref{fig:ideal-weld-evol-x80}).

\begin{figure}[H] 
    \centering
    \includegraphics[width=\textwidth]{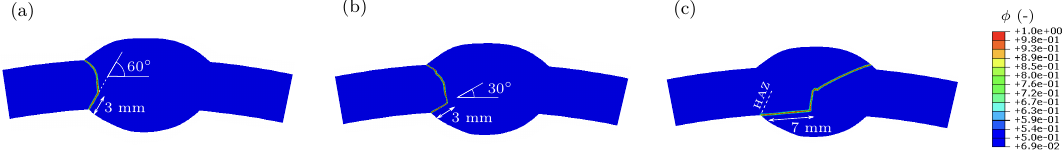}
    \caption{Investigating the role of external pre-existing defects on the failure of hydrogen transport pipelines. An initial flaw is assumed to be present at the weld-base material junction, near the HAZ region, and the influence of crack length and orientation (relative to the horizontal axis) is assessed by considering: (a) a 3 mm flaw with 60$^\circ$ orientation ($p_\mathrm{max} = 15$ MPa), (b) a 3 mm flaw with 30$^\circ$ orientation ($p_\mathrm{max} = 19$ MPa), and (c) a 7 mm flaw with 0$^\circ$ orientation ($p_\mathrm{max} = 28$ MPa). Results obtained for the two-pass weld considering X80 pipeline steel as the base and weld materials.}
    \label{fig:ideal-weld-flaw-haz}
\end{figure}

\begin{figure}[H] 
    \centering
   \includegraphics[width=0.95\textwidth]{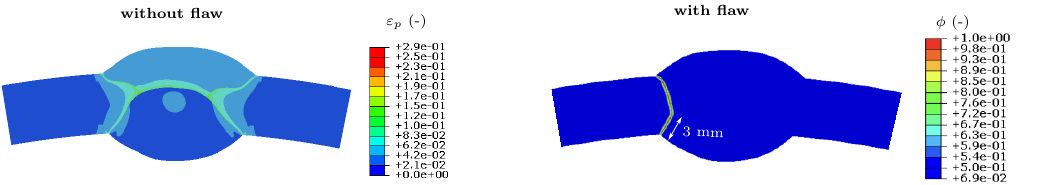}
    \caption{Investigating the role of an external pre-existing flaw that is located near the inner surface and aligned with the HAZ. Results obtained for the two-pass weld considering X52 pipeline steel as the base and weld materials. The finite element analysis reveals: a global yielding type of failure at $p_\mathrm{max} \approx 29$ MPa) in the absence of flaws (a), yet a brittle fracture event at a pressure of $p_\mathrm{max} = 19$ MPa in the presence of a 3 mm flaw along the HAZ (b).}
    \label{fig:ideal-weld-flaw-x52}
\end{figure}

Next, we studied the role of the crack orientation and length on performance. For this, we estimated the critical failure pressure for an existing flaw that was located near the HAZ, at the weld-base material junction, and which was inclined at 60$^\circ$, 30$^\circ$ and 0$^\circ$ from the horizontal axis. Two crack lengths were considered: 3 and 7 mm. As shown in \cref{fig:ideal-weld-flaw-haz}, the horizontal flaw exhibited a very different cracking pattern relative to the other two scenarios. Although no significant reduction in admissible pressure was predicted ($p_\mathrm{max} = 28$ MPa, relative to a defect-free estimate of $p_\mathrm{max} = 31$ MPa). However, the 3 mm cracks aligned closer to the HAZ appear to have a very significant impact on the pressure-carrying capacity. We find that pipeline failure occurs at $p_\mathrm{max} = 19$ MPa, for the 30$^\circ$ orientation, and at $p_\mathrm{max} = 15$ MPa, for the 60$^\circ$ orientation. 
Thus, defect location appears to play a critical role in fracture behaviour. Failure pressures of 15 MPa are of significant technological importance as they lie within the ranges considered for hydrogen transport. Furthermore, while this drop in load-carrying capacity is attained under specific conditions, these scenarios were selected to be relevant to existing pipelines. Indeed, defects of similar size and location have been reported on seam welds of gas transport pipelines and is expected that defects will originate near the inner surface and along the harder and more brittle HAZ region. When a similar analysis was conducted on the more ductile X52 pipeline steel, a change in failure mode from global yielding to cracking was observed. The predicted critical failure pressure dropped to $p_\mathrm{max} = 18$ MPa when a 3 mm flaw was present along the HAZ - see \cref{fig:ideal-weld-flaw-x52}. 

\subsection{Case study 2: three-pass seam weld}
\label{Sec:ThreePass}

\begin{figure}[H] 
    \centering
    \includegraphics[width=\textwidth]{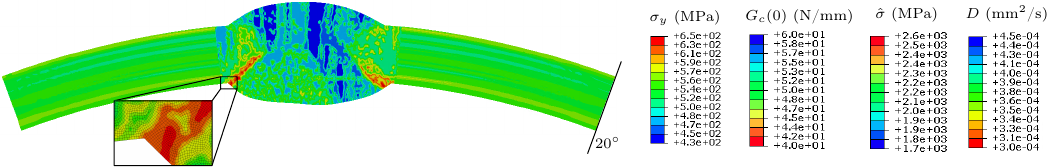}
    \caption{Spatial distribution of relevant material properties (yield stress, toughness, strength and diffusivity) as inferred from the microhardness map of the three-pass weld (see \cref{fig:ideal-weld-geo}). The results shown correspond to the X52 pipeline steel case.}
    \label{fig:bad-weld-mapping}
\end{figure}

A second seam weld configuration is investigated to assess the role of weld quality and geometry on performance. The second weld also facilitated the investigation of the influence of weld microstructural heterogeneity. As shown in \cref{fig:ideal-weld-geo}b, we considered a seam weld fabricated using three-passes. The overall weld dimensions were 21 x 13 mm. Microhardness mapping had been performed on this ex-service weld. This weld configuration is representative of seam welds that require a third, repair weld pass. The material properties adopted are largely those reported in \cref{tab:weld1-param} but some changes are introduced to incorporate the information provided by the microstructural hardness map. Namely, the ASTM-A370 relationship is adopted to quantitatively relate the HV measurements with yield strength, with $\sigma_y$ varying between 660 and 990 MPa for the X80 pipeline steel and between 430 and 645 MPa for the X52 pipeline steel (see \cref{fig:bad-weld-mapping}). The fracture-to-yield strength ratio was maintained at the previous level, and this brings a variation in the fracture strength ($\hat{\sigma}$) distribution, as shown in \cref{fig:bad-weld-mapping}. Also, since a higher strength typically corresponds to a lower toughness, the analysis performed considered that the fracture energy varied inversely to the microhardness. Finally, and consistent with experimental observations \cite{souzaEffectMicrostructureHydrogen2017}, the diffusivity was assumed to be lower in hard regions. The other parameters were kept the same as in \cref{tab:weld1-param}, with Poisson's ratio and strain hardening exponent being uniform across the section ($\nu=0.3$ and $n=0.1$, respectively). As observed in \cref{fig:bad-weld-mapping}, this microstructural mapping allows the variations in strength and toughness of the HAZ regions to be displayed.

\begin{figure}[H] 
    \centering
    \includegraphics[width=0.95\textwidth]{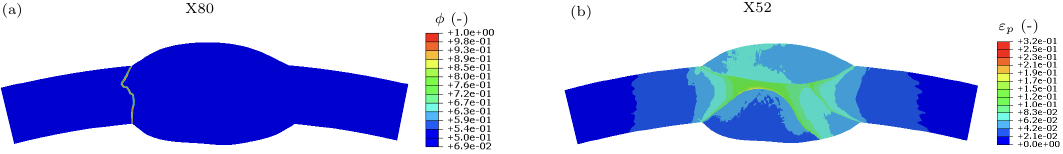}
    \caption{Computational predictions of weld structural integrity in hydrogen transport pipelines: (a) brittle fracture in an X80 pipeline at a critical pressure of $p_\mathrm{max} = 24$ MPa, (b) failure by global plastic yielding in an X52 pipeline.}
    \label{fig:bad-weld-failure-mode}
\end{figure}

\begin{figure}[H] 
    \centering
    \includegraphics[width=\textwidth]{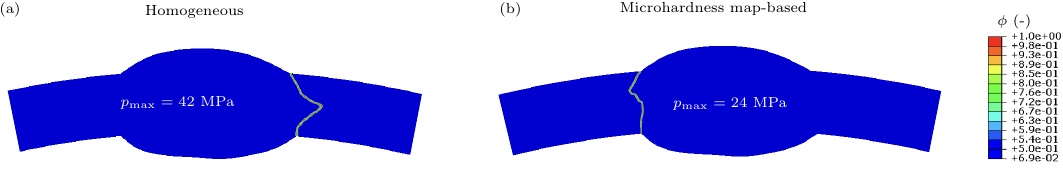}
    \caption{Investigating the role of accounting for the microstructural heterogeneity of seam welds. Cracking (phase field) contours for: (a) the case where material properties are considered to be homogeneous in the weld and base metal regions, and (b) the case where material properties are varied according to the microstructural hardness map (see Figs. \ref{fig:ideal-weld-geo} and \ref{fig:bad-weld-mapping}). Failure takes place due to cracking at critical pressures of $p_\mathrm{max} = 42$ MPa and $p_\mathrm{max} = 24$ MPa, respectively.}
    \label{fig:bad-weld-x80}
\end{figure}

As in the previous case study, a two-step analysis was conducted. This started with a welding process simulation to obtain the associated residual stresses, and was followed by a coupled deformation-diffusion-fracture analysis to determine the maximum admissible pressure. The evolution of the stress fields, the equivalent plastic distribution and the phase field fracture variability in this welded joint was qualitatively similar to that of the previous weld geometry despite the differences in the welding process and the heterogeneous distribution of material properties. 
However, interesting differences were found when inspecting the critical failure pressures and the cracking patterns. First, as shown in \cref{fig:bad-weld-failure-mode}, the X80 pipeline was predicted to fail due to cracking at a pressure of $p_\mathrm{max} = 24$ MPa, a lower magnitude than that reported for the two-pass weld in the absence of defects (31 MPa). The cracking pattern was also slightly different, with the crack nucleating at the intersection between the HAZ and the inner surface but growing mostly through the base metal. In contrast, the X52 pipeline was predicted to fail due to global plastic yielding, as observed in the two-pass weld.\\
Calculations were also conducted with homogeneous material properties, as in the two-pass weld case study, to quantify the influence of the inherent microstructural heterogeneity of welded components. The results, shown in \cref{fig:bad-weld-x80}, reveal a notable influence of the material microstructure. When the material properties were taken to be homogeneous in the base metal and the weld, then failure was predicted to take place at a pressure of $p_\mathrm{max} = 42$ MPa. However, when the microstructural heterogeneity of the material was accounted for, the predicted fracture pressure was reduced by approximately 50\%, to $p_\mathrm{max} = 24$ MPa. Interestingly, the failure pressure obtained for the homogeneous case study is significantly higher than that reported for the 17 x 14 mm two-pass weld under the same conditions (31 MPa). This suggests that the three-pass weld with dimensions 21 x 13 mm is more resistant to hydrogen-assisted fractures. This also suggests that the lower bound of $p_\mathrm{max} = 15$ MPa obtained for the two-pass weld in the presence of a HAZ-aligned 3 mm flaw could be further reduced if microstructural heterogeneities are accounted for.

\section{Conclusions}
\label{sec:conclusions}
We have presented a computational framework for predicting the maximum admissible pressure in hydrogen transport pipelines. The framework combines welding process modelling, to determine the residual stress state, with coupled deformation-diffusion-fracture (embrittlement) simulations, to quantify critical fracture pressures. {The focus is on maximising simplicity and predictive abilities. As such, the model relies only on parameters with a clear physical interpretation, that can be independently measured.} Virtual experiments were conducted on pipelines containing two types of seam welds. These were selected to represent real welds in gas transport pipelines: a two-pass weld of dimensions 17 x 14 mm, and a three-pass weld of dimensions 21 x 13 mm, for which a microstructural hardness map exists. The model was first benchmarked against crack growth resistance curves on X52 and X80 pipeline steels. Then the model was used to investigate the influence of residual stresses, welding quality and geometry, microstructure, internal and external defects, and hydrogen gas pressure. Key findings include:
\begin{itemize}
\item An accurate characterisation of residual stresses can be decisive in determining the failure mode. In the absence of pre-existing defects, X80 pipeline steels were predicted to fail by global plastic yielding (at $p_\mathrm{max} \approx 45$ MPa) if residual stresses are neglected. However,  cracks are found to nucleate and grow across the pipeline cross-section at 31 MPa H$_2$ pressures when the impact of residual stresses arising from the welding process was taken into account.
\item The presence of micrometer pores with volume fractions below 0.005 can reduce the admissible pressure in X80 pipelines to $p_\mathrm{max} = 24$ MPa. The presence of pores was not predicted to play a role in the performance of the more ductile X52 pipelines, which fail by global plastic yielding (in the absence of other defects).
\item The maximum admissible pressure was found to be very sensitive to the size, location and orientation of pre-existing defects. Those closer to the inner surface of the pipeline (where the hydrogen content is greater) and aligned with and near to the HAZ being the most harmful. 
\item The presence of 3 mm long surface flaws which are aligned with the HAZ can significantly reduce the pressure-carrying capacity of hydrogen transport pipelines. This realistic scenario results in a maximum admissible pressure of $p_\mathrm{max} = 15$ MPa for the X80 pipelines and of $p_\mathrm{max} = 19$ MPa for the X52 pipeline. 
\item The weld geometry and quality were found to be important. This highlights the need to characterise the welds of the existing natural gas pipeline infrastructure and to ensure that best practice is used when manufacturing new welds. 
\item The microstructural heterogeneity of the weld region is found to play a key role, potentially changing the critical failure pressure by approximately 50\% relative to the consideration of homogeneous weld and base materials.
\end{itemize}
 
These findings and the computational tools developed are expected to be of significant importance to the potential deployment of a hydrogen transport pipeline network, given the challenges associated with laboratory and field assessment.\\

{Multiple avenues of future work can be envisaged. First, significant work is ongoing in the weld process modelling community to develop more sophisticated descriptions of residual stresses and microstructural phases. One could build upon such modelling efforts not only to provide a more detailed description but also to validate with experimental testing (e.g., by comparing real and virtual hardness maps). In addition, other aspects of the deformation-diffusion-fracture modelling could be readily incorporated, if deemed relevant, from fatigue damage to explicit consideration of trapping sites.}

\section*{Declaration of Interest}
The authors declare that they have no known competing financial interests or personal relationships that could have appeared to influence the work reported in this paper.

\section*{Acknowledgements}

The authors acknowledge financial support from EPRI through the R\&D project ``\textit{Virtual Testing} of hydrogen-sensitive components''. E. Mart\'{\i}nez-Pa\~neda additionally acknowledges financial support from UKRI's Future Leaders Fellowship programme [grant MR/V024124/1]. T. K. Mandal gratefully acknowledges stimulating discussions on phase field modelling, image processing, and weld simulation with Adria Quintanas-Corominas (Barcelona Supercomputing Center, Spain), Abhinav Gupta (Vanderbilt University, USA), Yousef Navidtehrani (University of Oviedo, Spain) and Lucas Castro García (University of Oviedo, Spain). 

\appendix
\numberwithin{figure}{section}
\numberwithin{table}{section}
\renewcommand{\thefigure}{\Alph{section}.\arabic{figure}}
\renewcommand{\thetable}{\Alph{section}.\arabic{table}}

\section{Model verification: crack growth resistance curves}
\label{sec:Rcurve}
Here, we showcase the ability of the model to predict the crack growth resistance of pipeline steels and rationalise the values of the fracture parameters chosen in this study. For a set of given elastoplastic material properties, the crack growth resistance (R-curve) behaviour is primarily governed by the fracture energy $G_c$ and fracture strength $\hat{\sigma}$ of the material \cite{tvergaardRelationCrackGrowth1992,kristensen2020phase}. In the phase field model here adopted, the former is explicitly provided, as $G_c$ is an input variable, while the latter is defined through the choice of phase field length scale, as per \cref{eq:at2-l0}. It should be noted that for boundary value problems with sufficiently large cracks (longer than the transition flaw size), the fracture behaviour is toughness-dominated and crack growth is therefore governed by $G_c$ and independent of the fracture strength. However, $\hat{\sigma}$ does influence the amount of dissipation during subcritical crack growth \cite{tvergaardRelationCrackGrowth1992,kristensen2020phase}, and thus should be chosen appropriately. This R-curve modelling exercise allows for this. Another aspect to note is that while the $G \equiv J$ analogy holds, the critical value of the energy release rate in the phase field model, $G_c$, does not correspond to $J_{Ic}$, as the latter is measured after 0.2 mm of crack extension while $G_c$ corresponds to the onset of crack growth. As such, the measured $J_{Ic}$ value provides only one data point and the entire R-curve is needed to ensure that the model provides an accurate characterisation of crack growth resistance. An appropriate set of $(G_c, \hat{\sigma})$ values has to be determined, through appropriate choices of $G_c$ and $\ell$. In this regard, it should be noted that measuring $G_c$ (also referred to as $J_0$) is not straightforward, given the challenges associated with capturing the onset of crack growth (particularly in hydrogen-containing environments).\\

\begin{figure}[H] 
    \centering
    \includegraphics[width=0.8\textwidth]{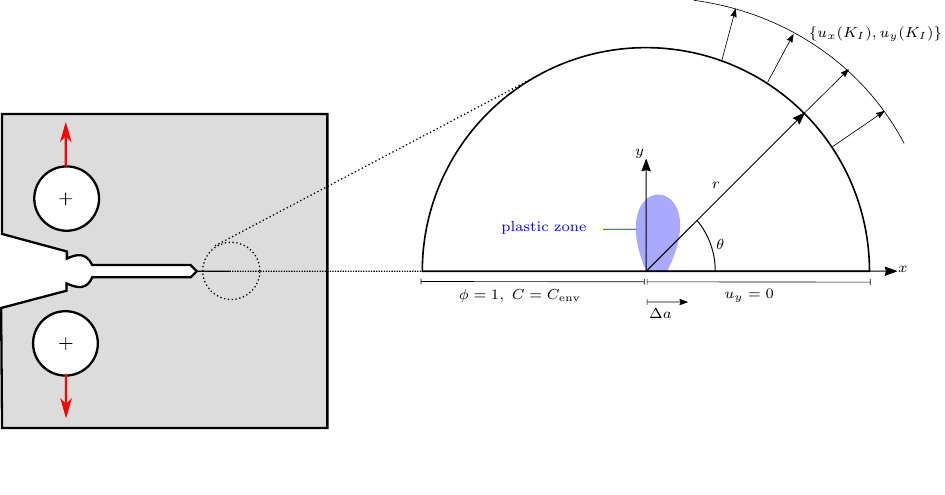}
    \caption{The boundary layer idealisation of a compact tension specimen under small-scale yield conditions. R-curves are obtained by applying a remote $K$ ($J$) and measuring the crack extension $\Delta a$.}
    \label{fig:BL-geo}
\end{figure}

\begin{figure}[H] 
    \centering
    \includegraphics[width=\textwidth]{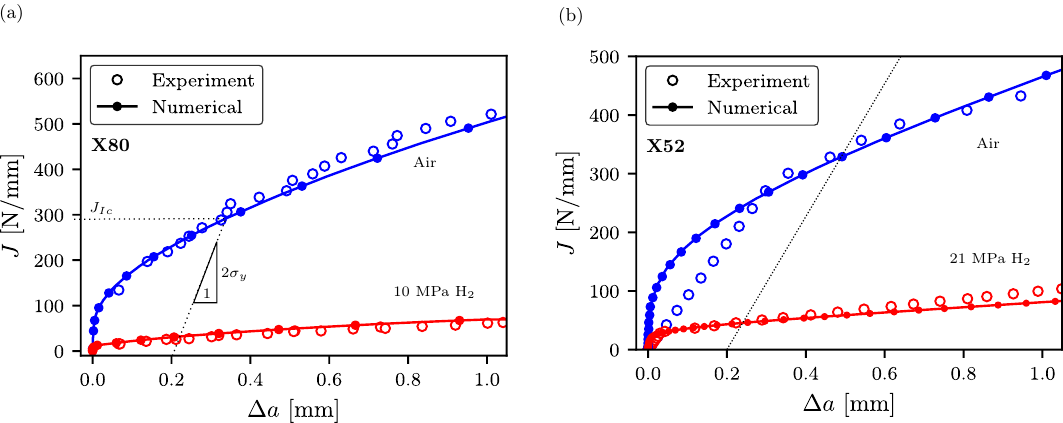}
    \caption{Predicting the crack growth resistance behaviour of (a) X80 pipeline steel \cite{shangDifferentEffectsPure2021}, and (b) X52 pipeline steel \cite{ronevichHydrogenassistedFractureResistance2021}, in both air and hydrogen environments.}
    \label{fig:Rcurve-calibration}
\end{figure}

Crack growth resistance curves are obtained by making use of the so-called boundary layer formulation, whereby William's solution \cite{williamsStressDistributionBase1957} for a remote elastic $K$-field is used to prescribe the displacement of the outer nodes (see \cref{fig:BL-geo}). Accordingly, the displacement field is given by,
\begin{equation}
u_i = \dfrac{K_I}{E}r^{1/2} f_i (\theta,\nu),
\end{equation}
where $r$ and $\theta$ are the coordinates of a polar coordinate system centred at the crack tip and $K_I$ is the mode I stress intensity factor, which under small scale yielding conditions is related to the $J$-integral as $J_I = K_I^2(1-\nu^2)/E$. As per William's solution \cite{williamsStressDistributionBase1957}, the function $f_i$ reads, 
\begin{align}
f_x = \dfrac{1+\nu}{\sqrt{2\pi}} \Big(3-4\nu-\cos{\theta}\Big)\cos{\left(\dfrac{\theta}{2}\right)}, \quad 
f_y = \dfrac{1+\nu}{\sqrt{2\pi}} \Big(3-4\nu-\cos{\theta}\Big)\sin{\left(\dfrac{\theta}{2}\right)}.
\end{align}
We take advantage of the reflective symmetry of the boundary value problem about the crack plane and simulate only half of the complete boundary layer, as shown in \cref{fig:BL-geo}. The outer radius is chosen to be sufficiently large so as to not influence the results: $R \gg R_p$, with $R_p$ being Irwin's estimate of the plastic zone length,
\begin{equation} 
R_p:= \frac{1}{3 \pi} \frac{K_I}{\sigma_y^2}.
\end{equation}
In all calculations, the characteristic element size is taken to be five times smaller than the phase field length scale, so as to ensure mesh objective results \cite{Mandal:EFM2019b}. The material properties for X80 and X52 pipeline steels are taken from the literature, and are consistent with the base metal parameters reported in the main text. Namely, the elastic properties are taken to be $E=187$ GPa and $\nu=0.3$ \cite{shangDifferentEffectsPure2021}, and the fit of the hardening law adopted to the uniaxial stress-strain data from Ref. \cite{caiExperimentalInvestigationHydrogen2022} renders a hardening exponent of $n=0.1$. The values of yield stress are $\sigma_y=660$ MPa and $\sigma_y=430$ MPa for, respectively, X80 and X52 pipeline steel (\cref{tab:database}). The numerical experiments mimic the conditions of the laboratory tests used as a benchmark (see Refs. \cite{shangDifferentEffectsPure2021,ronevichHydrogenassistedFractureResistance2021}). 
The hydrogen concentration corresponding to the reported H$_2$ pressure is determined using Sievert's law (\cref{eq:pressure}). Also, as in the experiments, a very slow loading rate ($\dot{K}_I \sim 0.05\;\mathrm{MPa}\;\sqrt{\mathrm{m}}\;/\mathrm{s}$) is adopted. The results obtained are shown in \cref{fig:Rcurve-calibration}. \\

As shown in \cref{fig:Rcurve-calibration}, for both X80 and X52 pipeline steels, the chosen values $G_c (0)$ and $\hat{\sigma}$ deliver a very good agreement with experimental measurements. These values are provided in \cref{tab:R-curve-param} and are subsequently used to predict the behaviour of welded components exposed to hydrogen environments (see \cref{sec:results}). It is important to note that hydrogen affects not only the load at which cracking initiates but also the slope of the R-curve, diminishing the degree of plastic dissipation. Hence, the hydrogen degradation law estimated from $J_{Ic}$ vs $C$ (or $p$) data is not necessarily the same as that describing the sensitivity of $G_c$ (or $J_0$) to $C$; i.e.,
\begin{align}
    \dfrac{G_{c}(C)}{J_{Ic}(C)} = d(C) \dfrac{G_{c}(0)}{J_{Ic} (0)}
\end{align}
where $d(C)\ge 1$ is a variable that characterises the higher degree of hydrogen degradation experienced by $J_{Ic}$ relative to $G_c$ (or $J_0$), as a result of the hindered plastic dissipation. Accordingly, \cref{eq:hydrogen-Gc} can be re-written as,
\begin{equation}
\label{eq:hydrogen-Gc-new}
G_{c}(C) = f(C)  d(C) G_{c}(0)
= \left[ \frac{G_{c}^{\text{min}}}{G_{c} (0)}  + \left( 1 - \dfrac{G_{c}^{\text{min}}}{G_{c}(0)} \right) \exp{ \left( - q_1 C^{q_2} \right)} \right] G_{c} (0).
\end{equation}

    \begin{table*}[htbp]
    \centering
    \setlength\fboxsep{0pt}
    \smallskip%
    \renewcommand\arraystretch{1}
    \begin{tabularx}{0.65\textwidth}{lcccc}
    \toprule
    Steel & $\sigma_y$ [MPa] & $G_{c}(0)$ [N/mm] & $G_{c}^\mathrm{min}$ [N/mm] & $\hat{\sigma}/\sigma_y$ [-]\\
    \midrule 
    X80  & 660 & 60 & 7 & 4 \\ 
    X52  & 430 & 60 & 16 & 4  \\ 
    \bottomrule
    \end{tabularx}
    \caption{Calibrated phase field model parameters for pipeline steels.}
    \label{tab:R-curve-param}
    \end{table*}




\begin{thebibliography}{10}
\expandafter\ifx\csname url\endcsname\relax
  \def\url#1{\texttt{#1}}\fi
\expandafter\ifx\csname urlprefix\endcsname\relax\def\urlprefix{URL }\fi
\expandafter\ifx\csname href\endcsname\relax
  \def\href#1#2{#2} \def\path#1{#1}\fi

\bibitem{lipiainenUseExistingGas2023}
S.~Lipi{\"a}inen, K.~Lipi{\"a}inen, A.~Ahola, E.~Vakkilainen, Use of existing
  gas infrastructure in {{European}} hydrogen economy, International Journal of
  Hydrogen Energy 48~(80) (2023) 31317--31329.

\bibitem{thawaniAssessingPressureLosses2023}
B.~Thawani, R.~Hazael, R.~Critchley, Assessing the pressure losses during
  hydrogen transport in the current natural gas infrastructure using numerical
  modelling, International Journal of Hydrogen Energy 48 (2023) 34463--34475.

\bibitem{laureysUseExistingSteel2022}
A.~Laureys, R.~Depraetere, M.~Cauwels, T.~Depover, S.~Hertel{\'e}, K.~Verbeken,
  Use of existing steel pipeline infrastructure for gaseous hydrogen storage
  and transport: {{A}} review of factors affecting hydrogen induced
  degradation, Journal of Natural Gas Science and Engineering 101 (2022)
  104534.

\bibitem{galyasEffectHydrogenBlending2023}
A.~B. Galyas, L.~Kis, L.~Tihanyi, I.~Szunyog, M.~Vadaszi, A.~Koncz, Effect of
  hydrogen blending on the energy capacity of natural gas transmission
  networks, International Journal of Hydrogen Energy 48~(39) (2023)
  14795--14807.

\bibitem{penevEconomicAnalysisHighpressure2019a}
M.~Penev, J.~Zuboy, C.~Hunter, Economic analysis of a high-pressure urban
  pipeline concept ({{HyLine}}) for delivering hydrogen to retail fueling
  stations, Transportation Research Part D: Transport and Environment 77 (2019)
  92--105.

\bibitem{rongTechnoeconomicAnalysisHydrogen2023}
Y.~Rong, S.~Chen, C.~Li, X.~Chen, L.~Xie, J.~Chen, R.~Long, Techno-economic
  analysis of hydrogen storage and transportation from hydrogen plant to
  terminal refueling station, International Journal of Hydrogen Energy 52
  (2023) 547--558.

\bibitem{Gangloff2003}
R.~P. Gangloff, {Hydrogen-assisted Cracking}, in: I.~Milne, R.~Ritchie,
  B.~Karihaloo (Eds.), Comprehensive Structural Integrity Vol. 6, Elsevier
  Science, New York, NY, 2003, pp. 31--101.

\bibitem{Gangloff2012}
R.~P. Gangloff, B.~P. Somerday, {Gaseous Hydrogen Embrittlement of Materials in
  Energy Technologies}, Woodhead Publishing Limited, Cambridge, 2012.

\bibitem{Djukic2019}
M.~B. Djukic, G.~M. Bakic, V.~{Sijacki Zeravcic}, A.~Sedmak, B.~Rajicic, {The
  synergistic action and interplay of hydrogen embrittlement mechanisms in
  steels and iron: Localized plasticity and decohesion}, Engineering Fracture
  Mechanics 216 (2019) 106528.

\bibitem{san2012technical}
C.~W. San~Marchi, B.~P. Somerday, Technical reference for hydrogen
  compatibility of materials., Tech. rep., {Sandia National Laboratories (SNL),
  Albuquerque, NM, and Livermore, CA} (2012).

\bibitem{boukorttHydrogenEmbrittlementEffect2018b}
H.~Boukortt, M.~Amara, M.~Hadj~Meliani, O.~Bouledroua, B.~Muthanna,
  R.~Suleiman, A.~Sorour, G.~Pluvinage, Hydrogen embrittlement effect on the
  structural integrity of {{API 5L X52}} steel pipeline, International Journal
  of Hydrogen Energy 43~(42) (2018) 19615--19624.

\bibitem{briottetInfluenceHydrogenOxygen2018}
L.~Briottet, H.~{Ez-Zaki}, Influence of {{Hydrogen}} and {{Oxygen Impurity
  Content}} in a {{Natural Gas}} / {{Hydrogen Blend}} on the {{Toughness}} of
  an {{API X70 Steel}}, in: {{ASME}} 2018 {{Pressure Vessels}} and {{Piping
  Conference}}, {American Society of Mechanical Engineers Digital Collection},
  2018, pp. 1--5.

\bibitem{ronevichEFFECTSTESTINGRATE2023}
J.~Ronevich, M.~Agnani, C.~S. Marchi, Effects of testing rate on
  hydrogen-assisted fracture of ferritic steels, in: International
  {{Conference}} on {{Fracture}}, {ICF15}, {Atlanta, USA}, 2023, pp. 1--2.

\bibitem{ronevichHydrogenAcceleratedFatigue2017}
J.~A. Ronevich, B.~P. Somerday, Z.~Feng, Hydrogen accelerated fatigue crack
  growth of friction stir welded {{X52}} steel pipe, International Journal of
  Hydrogen Energy 42~(7) (2017) 4259--4268.

\bibitem{ronevichFatigueCrackGrowth2020}
J.~A. Ronevich, E.~J. Song, Z.~Feng, Y.~Wang, C.~D'Elia, M.~R. Hill, Fatigue
  crack growth rates in high pressure hydrogen gas for multiple {{X100}}
  pipeline welds accounting for crack location and residual stress, Engineering
  Fracture Mechanics 228 (2020) 106846.

\bibitem{songNotchedtensilePropertiesHighpressure2017}
E.~J. Song, S.-W. Baek, S.~H. Nahm, U.~B. Baek, Notched-tensile properties
  under high-pressure gaseous hydrogen: {{Comparison}} of pipeline steel
  {{X70}} and austenitic stainless type {{304L}}, {{316L}} steels,
  International Journal of Hydrogen Energy 42~(12) (2017) 8075--8082.

\bibitem{mengHydrogenEffectsX802017}
B.~Meng, C.~Gu, L.~Zhang, C.~Zhou, X.~Li, Y.~Zhao, J.~Zheng, X.~Chen, Y.~Han,
  Hydrogen effects on {{X80}} pipeline steel in high-pressure natural
  gas/hydrogen mixtures, International Journal of Hydrogen Energy 42~(11)
  (2017) 7404--7412.

\bibitem{ishikawaIntegrityAssessmentLinepipes2022}
N.~Ishikawa, T.~Sakimoto, J.~Shimamura, J.~Wang, Y.-Y. Wang, Integrity
  {{Assessment}} of {{Linepipes}} for {{Transporting High Pressure Hydrogen
  Based}} on {{ASME B31}}.12, in: 2022 14th {{International Pipeline
  Conference}}, {American Society of Mechanical Engineers Digital Collection},
  2022, pp. 1--9.

\bibitem{kappesBlendingHydrogenExisting2023}
M.~A. Kappes, T.~E. Perez, Blending hydrogen in existing natural gas pipelines:
  Integrity consequences from a fitness for service perspective, Journal of
  Pipeline Science and Engineering (2023) 100141.

\bibitem{RILEM2021}
E.~Mart{\'{i}}nez-Pa{\~{n}}eda, {Progress and opportunities in modelling
  environmentally assisted cracking}, RILEM Technical Letters 6 (2021) 70--77.

\bibitem{Cheng2023}
G.~Cheng, X.~Wang, K.~Chen, Y.~Zhang, T.~A. Venkatesh, X.~Wang, Z.~Li, J.~Yang,
  {Probing the effects of hydrogen on the materials used for large-scale
  transport of hydrogen through multi-scale simulations}, Renewable and
  Sustainable Energy Reviews 182~(September 2022) (2023) 113353.

\bibitem{CS2020b}
E.~Mart{\'{i}}nez-Pa{\~{n}}eda, A.~D{\'{i}}az, L.~Wright, A.~Turnbull,
  {Generalised boundary conditions for hydrogen transport at crack tips},
  Corrosion Science 173 (2020) 108698.

\bibitem{Hageman2022}
T.~Hageman, E.~Mart{\'{i}}nez-Pa{\~{n}}eda, {An electro-chemo-mechanical
  framework for predicting hydrogen uptake in metals due to aqueous
  electrolytes}, Corrosion Science 208 (2022) 110681.

\bibitem{Sofronis1989}
P.~Sofronis, R.~M. McMeeking, {Numerical analysis of hydrogen transport near a
  blunting crack tip}, Journal of the Mechanics and Physics of Solids 37~(3)
  (1989) 317--350.

\bibitem{IJHE2016}
E.~Mart{\'{i}}nez-Pa{\~{n}}eda, S.~del Busto, C.~F. Niordson, C.~Beteg{\'{o}}n,
  {Strain gradient plasticity modeling of hydrogen diffusion to the crack tip},
  International Journal of Hydrogen Energy 41~(24) (2016) 10265--10274.

\bibitem{Dadfarnia2011}
M.~Dadfarnia, P.~Sofronis, T.~Neeraj, {Hydrogen interaction with multiple
  traps: Can it be used to mitigate embrittlement?}, International Journal of
  Hydrogen Energy 36~(16) (2011) 10141--10148.

\bibitem{AM2020}
R.~Fern{\'{a}}ndez-Sousa, C.~Beteg{\'{o}}n, E.~Mart{\'{i}}nez-Pa{\~{n}}eda,
  {Analysis of the influence of microstructural traps on hydrogen assisted
  fatigue}, Acta Materialia 199 (2020) 253--263.

\bibitem{IJP2021}
M.~Isfandbod, E.~Mart{\'{i}}nez-Pa{\~{n}}eda, {A mechanism-based multi-trap
  phase field model for hydrogen assisted fracture}, International Journal of
  Plasticity 144 (2021) 103044.

\bibitem{CupertinoMalheiros2022}
L.~{Cupertino Malheiros}, A.~Oudriss, S.~Cohendoz, J.~Bouhattate,
  F.~Th{\'{e}}bault, M.~Piette, X.~Feaugas, {Local fracture criterion for
  quasi-cleavage hydrogen-assisted cracking of tempered martensitic steels},
  Materials Science and Engineering: A 847 (2022) 143213.

\bibitem{Valverde-Gonzalez2022}
A.~Valverde-Gonz{\'{a}}lez, E.~Mart{\'{i}}nez-Pa{\~{n}}eda,
  A.~Quintanas-Corominas, J.~Reinoso, M.~Paggi, {Computational modelling of
  hydrogen assisted fracture in polycrystalline materials}, International
  Journal of Hydrogen Energy 47~(75) (2022) 32235--32251.

\bibitem{Serebrinsky2004}
S.~Serebrinsky, E.~A. Carter, M.~Ortiz, {A quantum-mechanically informed
  continuum model of hydrogen embrittlement}, Journal of the Mechanics and
  Physics of Solids 52~(10) (2004) 2403--2430.

\bibitem{Moriconi2014}
C.~Moriconi, G.~H{\'{e}}naff, D.~Halm, {Cohesive zone modeling of fatigue crack
  propagation assisted by gaseous hydrogen in metals}, International Journal of
  Fatigue 68 (2014) 56--66.

\bibitem{EFM2017}
S.~del Busto, C.~Beteg{\'{o}}n, E.~Mart{\'{i}}nez-Pa{\~{n}}eda, {A cohesive
  zone framework for environmentally assisted fatigue}, Engineering Fracture
  Mechanics 185 (2017) 210--226.

\bibitem{Elmukashfi2020}
E.~Elmukashfi, E.~Tarleton, A.~C.~F. Cocks, {A modelling framework for coupled
  hydrogen diffusion and mechanical behaviour of engineering components},
  Computational Mechanics 66 (2020) 189--220.

\bibitem{Colombo2020}
C.~Colombo, A.~{Zafra Garc{\'{i}}a}, J.~Belzunce, I.~{Fernandez Pariente},
  {Sensitivity to hydrogen embrittlement of AISI 4140 steel: A numerical study
  on fracture toughness}, Theoretical and Applied Fracture Mechanics 110 (2020)
  102810.

\bibitem{Dadfarnia2014}
M.~Dadfarnia, B.~P. Somerday, P.~E. Schembri, P.~Sofronis, J.~W. Foulk, K.~A.
  Nibur, D.~K. Balch, {On modeling hydrogen-induced crack propagation under
  sustained load}, Jom 66~(8) (2014) 1390--1398.

\bibitem{Anand2019}
L.~Anand, Y.~Mao, B.~Talamini, {On modeling fracture of ferritic steels due to
  hydrogen embrittlement}, Journal of the Mechanics and Physics of Solids 122
  (2019) 280--314.

\bibitem{Yu2019b}
H.~Yu, J.~S. Olsen, A.~Alvaro, L.~Qiao, J.~He, Z.~Zhang, {Hydrogen informed
  Gurson model for hydrogen embrittlement simulation}, Engineering Fracture
  Mechanics 217 (2019) 106542.

\bibitem{Depraetere2021}
R.~Depraetere, W.~{De Waele}, S.~Hertel{\'{e}}, {Fully-coupled continuum damage
  model for simulation of plasticity dominated hydrogen embrittlement
  mechanisms}, Computational Materials Science 200 (2021) 110857.

\bibitem{CMAME2018}
E.~Mart{\'{i}}nez-Pa{\~{n}}eda, A.~Golahmar, C.~F. Niordson, {A phase field
  formulation for hydrogen assisted cracking}, Computer Methods in Applied
  Mechanics and Engineering 342 (2018) 742--761.

\bibitem{Duda2018}
F.~P. Duda, A.~Ciarbonetti, S.~Toro, A.~E. Huespe, {A phase-field model for
  solute-assisted brittle fracture in elastic-plastic solids}, International
  Journal of Plasticity 102 (2018) 16--40.

\bibitem{Huang2020}
C.~Huang, X.~Gao, {Phase field modeling of hydrogen embrittlement},
  International Journal of Hydrogen Energy 45~(38) (2020) 20053--20068.

\bibitem{Wu2020b}
J.-Y. Wu, T.~K. Mandal, V.~P. Nguyen, {A phase-field regularized cohesive zone
  model for hydrogen assisted cracking}, Computer Methods in Applied Mechanics
  and Engineering 358 (2020) 112614.

\bibitem{Golahmar2022}
A.~Golahmar, P.~K. Kristensen, C.~F. Niordson, E.~Mart{\'{i}}nez-Pa{\~{n}}eda,
  {A phase field model for hydrogen-assisted fatigue}, International Journal of
  Fatigue 154 (2022) 106521.

\bibitem{Dinachandra2022}
M.~Dinachandra, A.~Alankar, {Adaptive finite element modeling of phase-field
  fracture driven by hydrogen embrittlement}, Computer Methods in Applied
  Mechanics and Engineering 391 (2022) 114509.

\bibitem{TAFM2020c}
P.~K. Kristensen, C.~F. Niordson, E.~Mart{\'{i}}nez-Pa{\~{n}}eda, {Applications
  of phase field fracture in modelling hydrogen assisted failures}, Theoretical
  and Applied Fracture Mechanics 110 (2020) 102837.

\bibitem{simo1998-computational}
J.~C. Simo, T.~J. Hughes, Computational Inelasticity, Vol.~7, {Springer Science
  \& Business Media}, 1998.

\bibitem{ASME_BPVC_IID}
{ASME Section II Part D}, {ASME} boiler and pressure vessel code, {ASME} Boiler
  and Pressure Vessel Committee on Materials, Three Park Avenue, {NY} (2010).

\bibitem{grifith1920phenomena}
A.~A. Griffith, The phenomena of rupture and flow in solids, Phil. Trans. R.
  Soc. Lond., A 221 (1920) 163.

\bibitem{Alvaro2015}
A.~Alvaro, I.~{Thue Jensen}, N.~Kheradmand, O.~M. L{\o}vvik, V.~Olden,
  {Hydrogen embrittlement in nickel, visited by first principles modeling,
  cohesive zone simulation and nanomechanical testing}, International Journal
  of Hydrogen Energy 40~(47) (2015) 16892--16900.

\bibitem{Francfort1998}
G.~A. Francfort, J.-J. Marigo, {Revisiting brittle fracture as an energy
  minimization problem}, Journal of the Mechanics and Physics of Solids 46~(8)
  (1998) 1319--1342.

\bibitem{Rabczuk:IAM2013a}
T.~Rabczuk, Computational methods for fracture in brittle and quasi-brittle
  solids: State-of-the-art review and future perspectives, ISRN Applied
  Mathematics (2013).

\bibitem{Bourdin2000}
B.~Bourdin, G.~A. Francfort, J.-J. Marigo, {Numerical experiments in revisited
  brittle fracture}, Journal of the Mechanics and Physics of Solids 48~(4)
  (2000) 797--826.

\bibitem{Miehe-1}
C.~Miehe, F.~Welschinger, M.~Hofacker, Thermodynamically consistent phase-field
  models of fracture: Variational principles and multi-field {FE}
  implementations, International Journal for Numerical Methods in Engineering
  83~(10) (2010) 1273--1311.

\bibitem{Chambolle2004}
A.~Chambolle, {An approximation result for special functions with bounded
  deformation}, Journal des Mathematiques Pures et Appliquees 83~(7) (2004)
  929--954.

\bibitem{taylorLatentEnergyRemaining1933}
G.~I. Taylor, H.~Quinney, The latent energy remaining in a metal after cold
  working, Proceedings of the Royal Society of London, Series A. 143~(849)
  (1933) 307--326.

\bibitem{Amor2009}
H.~Amor, J.~J. Marigo, C.~Maurini, {Regularized formulation of the variational
  brittle fracture with unilateral contact: Numerical experiments}, Journal of
  the Mechanics and Physics of Solids 57~(8) (2009) 1209--1229.

\bibitem{Ambati2016}
M.~Ambati, R.~Kruse, L.~De~Lorenzis, A phase-field model for ductile fracture
  at finite strains and its experimental verification, Computational Mechanics
  57~(1) (2016) 149--167.

\bibitem{Miehe-2}
C.~Miehe, M.~Hofacker, F.~Welschinger, A phase field model for rate-independent
  crack propagation: Robust algorithmic implementation based on operator
  splits, Computer Methods in Applied Mechanics and Engineering 199~(45-48)
  (2010) 2765 -- 2778.

\bibitem{PTRSA2021}
P.~K. Kristensen, C.~F. Niordson, E.~Mart{\'{i}}nez-Pa{\~{n}}eda, {An
  assessment of phase field fracture: crack initiation and growth},
  Philosophical Transactions of the Royal Society A: Mathematical, Physical and
  Engineering Sciences 379 (2021) 20210021.

\bibitem{andreasDiaz-comsol-pfm}
A.~D{\'i}az, J.~M. Alegre, I.~I. Cuesta, E.~{Mart{\'i}nez-Pa{\~n}eda}, A
  {COMSOL} framework for predicting hydrogen embrittlement - {Part II}: phase
  field fracture.  (in preparation).

\bibitem{Martinez-Paneda2020}
E.~Mart{\'{i}}nez-Pa{\~{n}}eda, {NEXTGEM - EPSRC New Investigator Award}, Tech.
  rep. (2020).

\bibitem{ronevichMaterialsCompatibilityConcerns2021}
J.~A. Ronevich, C.~San~Marchi, Materials {{Compatibility Concerns}} for
  {{Hydrogen Blended Into Natural Gas}}, in: {{ASME}} 2021 {{Pressure Vessels}}
  \& {{Piping Conference}}, {American Society of Mechanical Engineers Digital
  Collection}, 2021, pp. 1--6.

\bibitem{sanmarchiMaterialsEvaluationHydrogen2021}
C.~San~Marchi, J.~Ronevich, K.~Simmons, Materials {{Evaluation}} for {{Hydrogen
  Service}}., Tech. Rep. SAND2021-10712PE, {Sandia National Lab. (SNL-NM),
  Albuquerque, NM (United States); Sandia National Laboratories, SNL
  California} (Aug. 2021).

\bibitem{shangDifferentEffectsPure2021}
J.~Shang, J.~Z. Wang, W.~F. Chen, H.~T. Wei, J.~Y. Zheng, Z.~L. Hua, L.~Zhang,
  C.~H. Gu, Different effects of pure hydrogen vs. hydrogen/natural gas mixture
  on fracture toughness degradation of two carbon steels, Materials Letters 296
  (2021) 129924.

\bibitem{sanmarchiFractureFatigueCommercial2011}
C.~San~Marchi, B.~P. Somerday, K.~A. Nibur, D.~G. Stalheim, T.~Boggess,
  S.~Jansto, Fracture and {{Fatigue}} of {{Commercial Grade API Pipeline
  Steels}} in {{Gaseous Hydrogen}}, in: {{ASME}} 2010 {{Pressure Vessels}} and
  {{Piping Division}}/{{K-PVP Conference}}, {American Society of Mechanical
  Engineers Digital Collection}, 2011, pp. 939--948.

\bibitem{sanmarchiFractureResistanceFatigue2012}
C.~San~Marchi, B.~P. Somerday, K.~A. Nibur, D.~G. Stalheim, T.~Boggess,
  S.~Jansto, Fracture {{Resistance}} and {{Fatigue Crack Growth}} of {{X80
  Pipeline Steel}} in {{Gaseous Hydrogen}}, in: {{ASME}} 2011 {{Pressure
  Vessels}} and {{Piping Conference}}, {American Society of Mechanical
  Engineers Digital Collection}, 2012, pp. 841--849.

\bibitem{sanmarchiHYDROGENCOMPATIBILITYSTRUCTURAL2021}
C.~San~Marchi, R.~Shrestha, J.~Ronevich, hydrogen compatibility of structural
  materials in natural gas networks., Tech. Rep. SAND2021-8820C, {Sandia
  National Lab. (SNL-NM), Albuquerque, NM (United States)} (Jul. 2021).

\bibitem{martinHydrogenEmbrittlementFerritic2020}
M.~L. Martin, M.~J. Connolly, F.~W. DelRio, A.~J. Slifka, Hydrogen
  embrittlement in ferritic steels, Applied Physics Reviews 7~(4) (2020)
  041301.

\bibitem{ronevichHydrogenassistedFractureResistance2021}
J.~A. Ronevich, E.~J. Song, B.~P. Somerday, C.~W. San~Marchi, Hydrogen-assisted
  fracture resistance of pipeline welds in gaseous hydrogen, International
  Journal of Hydrogen Energy 46~(10) (2021) 7601--7614.

\bibitem{EJMAS2019b}
E.~Mart{\'{i}}nez-Pa{\~{n}}eda, S.~Fuentes-Alonso, C.~Beteg{\'{o}}n,
  {Gradient-enhanced statistical analysis of cleavage fracture}, European
  Journal of Mechanics - A/Solids 77 (2019) 103785.

\bibitem{ahmedNumericalSimulationSteel2019}
B.~Ahmed, O.~C.~E. Bahri, B.~Benattou, T.~Malika, Numerical {{Simulation}} of a
  {{Steel Weld Joint}} and {{Fracture Mechanics Study}} of a {{Compact Tension
  Specimen}} for {{Zones}} of {{Weld Joint}}, Frattura ed Integrit\`a
  Strutturale 13~(47) (2019) 17--29.

\bibitem{navidtehraniSimpleRobustAbaqus2021}
Y.~Navidtehrani, C.~Beteg{\'o}n, E.~{Mart{\'i}nez-Pa{\~n}eda}, A simple and
  robust {{Abaqus}} implementation of the phase field fracture method,
  Applications in Engineering Science 6 (2021) 100050.

\bibitem{Materials2021}
Y.~Navidtehrani, C.~Beteg{\'{o}}n, E.~Mart{\'{i}}nez-Pa{\~{n}}eda, {A unified
  Abaqus implementation of the phase field fracture method using only a user
  material subroutine}, Materials 14~(8) (2021) 1913.

\bibitem{IJSS2015}
E.~Mart{\'{i}}nez-Pa{\~{n}}eda, C.~Beteg{\'{o}}n, {Modeling damage and fracture
  within strain-gradient plasticity}, International Journal of Solids and
  Structures 59 (2015) 208--215.

\bibitem{pfczm-hydrogen2}
T.~K. Mandal, V.~P. Nguyen, J.-Y. Wu, Comparative study of phase-field damage
  models for hydrogen assisted cracking, Theoretical and Applied Fracture
  Mechanics 111 (2021) 102840.

\bibitem{arpin2003material}
K.~Arpin, T.~Trimble, Material properties test to determine ultimate strain and
  true stress-true strain curves for high yield steels, Tech. rep., Lockheed
  Martin, Inc., Schenectady, NY (US) (2003).

\bibitem{souzaEffectMicrostructureHydrogen2017}
R.~C. Souza, L.~R. Pereira, L.~M. Starling, J.~N. Pereira, T.~A. Sim{\~o}es,
  J.~a. C.~P. Gomes, A.~H.~S. Bueno, Effect of {{Microstructure}} on {{Hydrogen
  Diffusion}} in {{Weld}} and {{API X52 Pipeline Steel Base Metals}} under
  {{Cathodic Protection}}, International Journal of Corrosion 2017 (2017)
  e4927210.

\bibitem{timoshenko1959-plate-shell}
S.~Timoshenko, S.~{Woinowsky-Krieger}, et~al., Theory of Plates and Shells,
  Vol.~2, {McGraw-hill New York}, 1959.

\bibitem{tvergaardRelationCrackGrowth1992}
V.~Tvergaard, J.~W. Hutchinson, The relation between crack growth resistance
  and fracture process parameters in elastic-plastic solids, Journal of the
  Mechanics and Physics of Solids 40~(6) (1992) 1377--1397.

\bibitem{kristensen2020phase}
P.~K. Kristensen, C.~F. Niordson, E.~{Mart{\'i}nez-Pa{\~n}eda}, A phase field
  model for elastic-gradient-plastic solids undergoing hydrogen embrittlement,
  Journal of the Mechanics and Physics of Solids 143 (2020) 104093.

\bibitem{williamsStressDistributionBase1957}
M.~L. Williams, On the {{Stress Distribution}} at the {{Base}} of a
  {{Stationary Crack}}, Journal of Applied Mechanics 24~(1) (1957) 109--114.

\bibitem{Mandal:EFM2019b}
T.~K. Mandal, V.~P. Nguyen, J.-Y. Wu, Length scale and mesh bias sensitivity of
  phase-field models for brittle and cohesive fracture, Engineering Fracture
  Mechanics 217~(106532) (2019).

\bibitem{caiExperimentalInvestigationHydrogen2022}
L.~Cai, G.~Bai, X.~Gao, Y.~Li, Y.~Hou, Experimental investigation on the
  hydrogen embrittlement characteristics and mechanism of natural gas-hydrogen
  transportation pipeline steels, Materials Research Express 9~(4) (2022)
  046512.

\end{thebibliography}
\end{document}